\documentclass[aps,pra,twocolumn,superscriptaddress]{revtex4}
\usepackage{amsfonts,amssymb,graphicx,amsmath}
\usepackage{epstopdf}
\usepackage{amsmath}
\usepackage[usenames,dvipsnames]{xcolor}
\usepackage{pdfpages}
\usepackage{footmisc}
\usepackage[normalem]{ulem}
\usepackage{array}
\usepackage{color}

\begin{document}
\title{Exact momentum-space analysis of small spin-1/2 $J_1$-$J_2$ rings}
\author{Zimeng Li}
\affiliation{Center for Quantum Technology Research, and Key Laboratory of Advanced Optoelectronic Quantum Architecture and Measurements (MOE), School of Physics, Beijing Institute of Technology, Beijing 100081, China}
\author{Ning Wu}
\email{wunwyz@gmail.com}
\affiliation{Center for Quantum Technology Research, and Key Laboratory of Advanced Optoelectronic Quantum Architecture and Measurements (MOE), School of Physics, Beijing Institute of Technology, Beijing 100081, China}
\begin{abstract}
This paper considers an $N$-site spin-1/2 $J_1$-$J_2$ ring with $N=6$ and $8$. With the help of a set of exact few-magnon Bloch states, we obtain the block-diagonalized Hamiltonian consisting of block matrices of at most four dimensions. Partial of the eigenstates are analytically solved. For the six-site anisotropic ring, we reveal a subset of eigenstates that are simultaneous eigenstates of the Hamiltonian and the total angular momentum operator, even though the latter is not conserved. For both the six- and eight-site isotropic rings, we achieve momentum-space manifestations of several important states, including the famous Majumdar-Ghosh (MG) ground states and the Hamada-Kane-Nakagawa-Natsume (HKNN) ground state. The equivalence of these states with their real-space counterparts is explicitly shown for $N=6$. The structure of the HKNN ground state for small rings suggests that for any even number $N$ this state might behave like a ``bound state" with $N/2$ successive down spins binding together. 
\end{abstract}
\maketitle
\section{Introduction}
\par Strongly correlated systems with competing interactions provide an interesting and important platform to study the interplay between strong correlation, quantum fluctuation, and frustration. Among these, quantum spin chains with both nearest-neighbor (NN) and next-nearest-neighbor (NNN) exchange interactions (or the $J_1$-$J_2$ chain) have been long studied~\cite{MG,Niemeijer1970,Ono1972,Klein1976,Bader1979,Haldane1982,Hamada1988,TonegawaAFM,Tonegawa1989,Okamoto1992,Krivnov1996,Dmitriev2006,Meisner2006,Furusaki2008,Sudan2009,Furusaki2010,Furusaki2012,Kumar2015,Kumar2016,Kumar2017,Kumar2025,AJP,PRB2024,Physica2025} and attract recent attentions due to its relevance to quantum simulations in cold atom systems.
\par Unlike the spin-1/2 nearest-neighbor XXZ chain that can be solved exactly via the Bethe ansatz~\cite{Bethe,Hulthen}, the $J_1$-$J_2$ chain is not integrable and one has to resort to advanced numerical or field-theoretical methods to study large chains. Even in the case of the antiferromagnetic XXX chain, the Bethe ansatz generally cannot provide satisfactory descriptions of all the eigenstates. This stimulated the emergence of several works in the 1960's on short chains in which the magnon dispersion, the low-energy excitations, and thermodynamical properties could be accessed~\cite{Orbach1959,Mattheiss1961,Cloizeaux1962,Bonner1964}. In this sense, studying short $J_1$-$J_2$ chains is of theoretical interest. 
\par In a seminal paper by Majumdar and Ghosh~\cite{MG}, the isotropic $J_1$-$J_2$ chain of $N=4,6$, and $8$ sites are block diagonalized by using a set of basis functions constructed by Hulth\'en in the study of the XXX chain~\cite{Hulthen}. However, due to the lack of translational invariance, Hulth\'en's basis functions are unable to realize a further decomposition of the fourteen-dimensional zero-angular-momentum subspace of the eight-site $J_1$-$J_2$ ring~\cite{MG}. A more natural choice of the basis states are the translationally invariant Bloch states, which are relevant to the study of few-magnon excitations upon the fully polarized state~\cite{PRB2024,Physica2025,Majutwomagnon,Ono1970,Loly1986,Chubukov,Kuzian,Kecke}. Recently, a set of exact two- and three-magnon Bloch states for finite-size chains are constructed~\cite{PRB2022} and are subsequently employed to study the formation of two-magnon bound states in the higher-spin ferromagnetic XXZ chain~\cite{PRBL2024} and $J_1$-$J_2$ chain~\cite{PRB2024,Physica2025}. The Bloch-state method achieves a block diagonalization of the Hamiltonian in each momentum sector and facilitates the analytical or semianalytical treatment of the $n$-magnon problem with $n\leq 3$. The exact Bloch states for $n\geq4$ and arbitrary $N$ are still missing in the literature due to their complexity.
\par So far a lot of research efforts have been devoted to the ground-state properties of $J_1$-$J_2$ chains under general parameters (see a recent review and references therein~\cite{Kumar2025}). Due to the lack of integrability of the model, most of these properties for large systems can only be obtained by numerical methods such as exact diagonalization or density matrix renormalization group. Nevertheless, for some particular system parameters the ground-state wave function can be written down in a closed form. Two well-known examples are the antiferromagnetic-antiferromagnetic (AF-AF) Majumdar-Ghosh (MG) chain with $J_1=2J_2<0$~\cite{MG} and the ferromagnetic-antiferromagnetic (FM-AF) Hamada-Kane-Nakagawa-Natsume (HKNN) chain with $J_1=-4J_2>0$~\cite{Hamada1988}. Both the twofold MG ground states and the HKNN ground state are constructed based on the splitting of the total Hamiltonian into local ones, and then by simultaneously minimizing them. As a result, they are usually linear combinations of various dimer states and may have complicated structures in the real space. On the other hand, for periodic chains these states (or their linear superpositions) are also eigenstates of the translation operator and have definite momentum. Therefore it would be interesting to reveal the correspondence between the real-space and momentum-space representations of the above states, even though it is generally not straightforward to do so for large rings.
\par In this work, we use the aforementioned few-magnon Bloch states to achieve block diagonalizations of a six-site anisotropic $J_1$-$J_2$ ring [with Hamiltonian $H(6)$] and an eight-site isotropic ring [with Hamiltonian $H^{(\mathrm{isp})}(8)$], hence providing an exact momentum-space analysis of these small rings. Along this line we note that a four-site isotropic ring $H^{(\mathrm{isp})}(4)$ has been recently discussed~\cite{AJP}. With the help of good quantum numbers such as the number of magnons $n$, the wave number $k$, or the total angular momentum $l$, both $H(6)$ and $H^{(\mathrm{isp})}(8)$ are block diagonalized into smaller matrices whose dimensions are no more than four. Even though the total angular momentum is not conserved in an anisotropic ring, it is found that there exist simultaneous eigenstates of $H(6)$ and the total angular momentum operator. In the isotropic case, all the block matrices of $H^{(\mathrm{isp})}(6)$ can be solved in terms of linear functions or square roots. In general, the block matrices for $H^{(\mathrm{isp})}(8)$ cannot be analytically solved since cubic or quartic equations are involved. Interestingly, we find that some of these matrices admit analytical solutions at several particular values of the ratio $J_1/J_2$, including both the MG and HKNN points. Consequently, for both $H^{(\mathrm{isp})}(6)$ and $H^{(\mathrm{isp})}(8)$ we are able to obtain analytical expressions for the MG and HKNN ground-state wave functions in the Bloch basis, which are explicitly shown to be equivalent to their real-space counterparts in the case of $N=6$.  
\par We calculate the full spectrum of $H^{(\mathrm{isp})}(6)$ and $H^{(\mathrm{isp})}(8)$, which consists of 14 and 44 distinct energy levels, respectively. It is known that in the limiting case of an antiferromagnetic XXX ring ($J_1/|J_2|\to-\infty$) the two-spinon excitation upon the singlet ground state has spin 1~\cite{Cloizeaux1962,FT81}. We explore how these lowest excited states for fixed $k$ evolve with varying $J_1/|J_2|$. We also calculate the weight of each Bloch state and the spin-spin correlations in the ground state of $H^{(\mathrm{isp})}(8)$, with the latter consistent with previous results~\cite{TonegawaAFM,Tonegawa1989} obtained by exact diagonalization. In particular, we find that for $N=6$ ($N=8$) the Bloch state with the down spins lying on three (four) successive sites has the largest weight in the HKNN ground state, indicating that this state behaves like a $N/2$-magnon ``bound state". We conjecture that this property holds for arbitrary even $N$.  
\par It should be mentioned that this work does not aim to make a detailed analysis of ground-state properties of large $J_1$-$J_2$ rings since many related works have been done. Instead, our purpose is to treat the $J_1$-$J_2$ ring as analytical as possible so as to build up a connection between real-space and momentum-space descriptions of the model. This forces us to work with small rings. The rest of the paper is organized as follows. In Sec.~\ref{SecII} we introduce the $J_1$-$J_2$ ring and review some known results about the model that are valid for arbitrary system size. In Sec.~\ref{SecIII} and \ref{SecIV} we provide detailed momentum-space analysis of  $H(6)$ and $H^{(\mathrm{isp})}(8)$, respectively. Conclusions are drawn in Sec.~\ref{SecV}.
\section{Model, conserved quantities, and some known results}\label{SecII}
\par For the sake of readability, we introduce the $N$-site $J_1$-$J_2$ ring and review some known results about this model. All the results presented in this section are valid for an arbitrary even integer $N$.
\subsection{The $J_1$-$J_2$ ring and conserved quantities}\label{SecII1}
\par We consider the $N$-site $J_1$-$J_2$ ring described by the Hamiltonian
\begin{eqnarray}\label{Haml}
H(N)&=&H_{\mathrm{NN}}+H_{\mathrm{NNN}},\nonumber\\
H_{\mathrm{NN}}&=&- J_1\sum^{N}_{j=1} (S^x_{j}S^x_{j+1}+S^y_{j}S^y_{j+1}+\Delta_1 S^z_{j}S^z_{j+1})\nonumber\\
H_{\mathrm{NNN}}&=&- J_2\sum^{N}_{j=1} (S^x_{j}S^x_{j+2}+S^y_{j}S^y_{j+2}+\Delta_2 S^z_{j}S^z_{j+2}),
\end{eqnarray}
where $\vec{S}_j=(S^x_j,S^y_j,S^z_j)$ are spin-1/2 operators on site $j$, $J_1$ and $J_2$ are respectively the exchange interactions between two NN and NNN spins with $\Delta_1, \Delta_2>0$ the corresponding interaction anisotropies. We assume even $N$ and impose periodic boundary conditions $\vec{S}_j=\vec{S}_{N+j}$, so that the chain has translational invariance. We denote by $H^{(\mathrm{isp})}(N)$ the Hamiltonian in the isotropic case with $\Delta_1=\Delta_2=1$. The fully polarized state $|F\rangle=|\uparrow\ldots\uparrow\rangle$ is an obvious eigenstate of $H$ with eigenenergy $E_F=-NJ_+/2$, where $J_\pm\equiv (J_1\Delta_1\pm J_2\Delta_2)/2$.
\par The Hamiltonian $H$ has two obvious conserved quantities, the total magnetization $M=\sum^N_{j=1}S^z_j$ and the translation operator $T$ defined by $T|j_1,\ldots,j_n\rangle=|j_1+1,\ldots,j_n+1\rangle$, where $|j_1,\ldots,j_n\rangle\equiv S^-_{j_1}\cdots S^-_{j_n}|F\rangle$ is a general ``$n$-magnon" real-space state in the magnetization sector with $M=N/2-n$. Here, the distinct site indices $j_i$'s satisfy $1\leq j_1<j_2<\ldots<j_n\leq N$. An equivalent definition of $T$ is given by $TS^-_jT^{-1}=S^-_{j+1}$. It is apparent that $T^N=1$, giving the $N$ distinct eigenvalues of $T$, $t_m=e^{i2\pi m/N}$ $(m=1,2,\ldots,N)$. A general state $|\phi\rangle$ is said to carry momentum $k$ if it satisfies $T|\phi\rangle=e^{-ik}|\phi\rangle$. For example, $|F\rangle$ carries zero momentum since $T|F\rangle=|F\rangle$. Although $|F\rangle$ is not always the ground state of $H$, we will take it as a reference state with respect to which the $n$-magnon sector is defined. The existence of the above two conserved quantities indicates that any eigenenergy $E_n(k)$ and eigenstate $|\psi_n(k)\rangle$ can be labeled by the two quantum numbers $n$ and $k$. The model is also invariant under the $\mathbb{Z}_2$ transformation $\vec{S}_j\to-\vec{S}_j$ and the reflection transformation  $\vec{S}_j\to\vec{S}_{N-j}$. As a result, $E_n(k)$ has the symmetry properties $E_n(k)=E_{N-n}(k)$ and $E_n(k)=E_n(-k)$. We can thus restrict ourselves to the sectors with $0\leq n\leq N/2$ and $-\pi\leq k\leq 0$ when the energy spectrum is merely concerned.
\par In the isotropic case with $\Delta_1=\Delta_2=1$, the total angular momentum $\vec{L}^2$ of all the spins is further conserved, i.e., $[H^{(\mathrm{isp})},\vec{L}^2]=0$, where $\vec{L}\equiv \sum^N_{j=1}\vec{S}_j$. The addition of the $N$ spins-1/2 results in the total angular momenta $l=N/2,N/2-1,\ldots,0$. As shown by Dicke~\cite{Dicke}, the angular momentum $l$ appears
\begin{eqnarray}\label{dNl}
d(N,l)=\binom{N}{l+N/2}-\binom{N}{l+N/2+1}
\end{eqnarray}
times. Any eigenvalue (eigenstate) of $H^{(\mathrm{isp})}$ can be labelled by $n$, $k$, and $l$ as $E_{n,l}(k)$ ($|\psi_{n,l}(k)\rangle$). By expanding $\vec{L}^2$ as $\vec{L}^2=\frac{3N}{4}+2\sum_{i<j}\vec{S}_i\cdot \vec{S}_j$, we also have $[\vec{L}^2,T]=0$.
\par The one-magnon sector is easy to analyze. One can construct $N$ one-magnon Bloch states $
|\xi(k)\rangle=\frac{1}{\sqrt{N}}\sum^{N-1}_{j=0}e^{ikj}T^j|1\rangle$, where $k\in K_1=\left\{-\pi,-\pi+\frac{2\pi}{N},\ldots,0,\ldots,\pi-\frac{2\pi}{N}\right\}$. It is easy to see that $T|\xi(k)\rangle=e^{-ik}|\xi(k)\rangle$, i.e., $|\xi(k)\rangle$ carries momentum $k$. Since $[H,T]=0$ and $\{|\xi(k)\rangle\}$ are eigenstates of $T$ with distinct eigenvalues $\{e^{-ik}\}$, the state $|\xi(k)\rangle$ must also be an eigenstate of $H$, $H|\xi(k)\rangle=E_1(k)|\xi(k)\rangle$, where $E_1(k)=- (J_1\cos k +J_2\cos 2k) -J_+$ is the corresponding eigenenergy. The one-magnon problem also admit analytical treatment in certain long-range spin models~\cite{Feldman}.
\par Similarly, from $[\vec{L}^2,T]=0$ we assert that $|\xi(k)\rangle$ is also an eigenstate of $\vec{L}^2$ for any $\Delta_1$ and $\Delta_2$, although $\vec{L}^2$ is generally not conserved. A simultaneous eigenstate of $H$ and $\vec{L}^2$ under the condition $[H,\vec{L}^2]\neq 0$ will be referred to as a type-HL state. As we will see, type-HL states also exist in other magnetization sectors.
\subsection{Some known results}
\par We now review some known results about this model. We first discuss a class of eigenstates called ``zero-energy states" (ZESs)~\cite{PRB2022,PRB2024}, which are eigenstates degenerate with the ferromagnetic state $|F\rangle$. Let us define the collective lowering operator $\mathcal{L}_k\equiv\sum^N_{j=1}e^{ikj}S^-_j$ and suppose $k^*$ is any wave number belonging to the set $K_1$. It is shown in Ref.~\cite{PRB2024} that for arbitrary $J_1$ and $J_2$ the state $(\mathcal{L}_{k^*})^n|F\rangle$ is a ZES in the $n$-magnon sector of $H$, provided the conditions
\begin{eqnarray}\label{cond}
\Delta_1=\cos k^*,~\Delta_2=\cos 2k^*
\end{eqnarray}
are satisfied. From $T(\mathcal{L}_{k^*})^n|F\rangle=T(\mathcal{L}_{k^*})^nT^{-1}T|F\rangle=e^{-ink^*}(\mathcal{L}_{k^*})^n|F\rangle$ we see that $(\mathcal{L}_{k^*})^n|F\rangle$ carries momentum $nk^*$. It is interesting to note that similar structures as the ZESs are recently observed in interacting bosonic systems~\cite{Song}. Since $k^*=0\in K_1$, the isotropic model $H^{(\mathrm{isp})}$ has at least one zero-momentum ZES $(\mathcal{L}_{0})^n|F\rangle$ in each $n$-magnon sector, and hence the ferromagnetic state $|F\rangle$ is at least $(N+1)$-fold degenerate. In addition, these $N+1$ ZESs all belong to the total angular momentum $l=N/2$ since $[\vec{L}^2,\mathcal{L}_0]=0$.
\par The ground-state properties of $H$ or $H^{(\mathrm{isp})}$ depend on the signs of $J_1$ and $J_2$:
\par (i) $J_1>0$ and $J_2>0$ (FM-FM).
\par The ground state of $H$ is the ferromagnetic state $|F\rangle$ for $\Delta_1\geq 1$ and $\Delta_2\geq 1$.
\par (ii) $J_1<0$ and $J_2>0$ (AF-FM).
\par According to a theorem of Lieb and Mattis~\cite{Lieb}, the ground state of $H^{(\mathrm{isp})}$ in this case belongs to $l=0$.
\par (iii) $J_1<0$ and $J_2<0$ (AF-AF).
\par The ground state of the isotropic Hamiltonian $H^{(\mathrm{isp})}$ in this case is well studied and there exists a critical point $J_1/|J_2|\approx- 4.148$ separating a dimer phase and a spin fluid phase~\cite{Haldane1982,TonegawaAFM,Okamoto1992,Kumar2016,Kumar2025} in the thermodynamic limit. In particular, Majumdar and Ghosh find at the special point $J_1/J_2=2$ that the ground state can be written down explicitly and is twofold degenerate~\cite{MG}:
\begin{eqnarray}\label{MG_GS}
|\psi^{(1)}_{\mathrm{MG}}\rangle&=&\left(\frac{1}{\sqrt{2}}\right)^{N/2}[12][34]\cdots[N-1,N],\nonumber\\
|\psi^{(2)}_{\mathrm{MG}}\rangle&=&\left(\frac{1}{\sqrt{2}}\right)^{N/2}[23][45]\cdots[N,1],
\end{eqnarray}
where $[i,j]\equiv |\uparrow\rangle_i|\downarrow\rangle_j-|\downarrow\rangle_i|\uparrow\rangle_j$ is a singlet state on sites $i$ and $j$. These two dimer states (or their linear superpositions) are also the only ground states at the MG point with ground-state energy $E_{\mathrm{MG}}=3NJ_1/8$~\cite{Caspers}. Either $|\psi^{(1)}_{\mathrm{MG}}\rangle$ or $|\psi^{(2)}_{\mathrm{MG}}\rangle$ is constructed in the real space as the simultaneous ground state of a set of local Hamiltonians~\cite{MG} and is a linear combination of $2^{N/2}$ different Ising configurations. However, $|\psi^{(1)}_{\mathrm{MG}}\rangle$ and $|\psi^{(2)}_{\mathrm{MG}}\rangle$ are not orthogonal since their expansions share certain common Ising configurations. Actually, the scalar product of the two is $\langle \psi^{(2)}_{\mathrm{MG}}|\psi^{(1)}_{\mathrm{MG}}\rangle=2(-1/2)^{N/2}$~\cite{TonegawaAFM,Tasaki}. Note that $T|\psi^{(1)}_{\mathrm{MG}}\rangle=|\psi^{(2)}_{\mathrm{MG}}\rangle$ and $T|\psi^{(2)}_{\mathrm{MG}}\rangle=|\psi^{(1)}_{\mathrm{MG}}\rangle$, we can construct the following two normalized orthogonal ground states
\begin{eqnarray}\label{MG_GSeo}
|\psi^{(\mathrm{e})}_{\mathrm{MG}}\rangle&=&\frac{1}{\sqrt{2+4(-1/2)^{N/2}}}(|\psi^{(1)}_{\mathrm{MG}}\rangle+|\psi^{(2)}_{\mathrm{MG}}\rangle),\nonumber\\
|\psi^{(\mathrm{o})}_{\mathrm{MG}}\rangle&=&\frac{1}{\sqrt{2-4(-1/2)^{N/2}}}(|\psi^{(1)}_{\mathrm{MG}}\rangle-|\psi^{(2)}_{\mathrm{MG}}\rangle),
\end{eqnarray}
which satisfy $T|\psi^{(\mathrm{e})}_{\mathrm{MG}}\rangle=|\psi^{(\mathrm{e})}_{\mathrm{MG}}\rangle$ and $T|\psi^{(\mathrm{o})}_{\mathrm{MG}}\rangle=-|\psi^{(\mathrm{o})}_{\mathrm{MG}}\rangle$, indicating that $|\psi^{(\mathrm{e})}_{\mathrm{MG}}\rangle$ ($|\psi^{(\mathrm{o})}_{\mathrm{MG}}\rangle$) lies in the $k=0$ ($k=-\pi$) subspace of the zero-magnetization (with $n=N/2$) sector.
\par (iv) $J_1>0$ and $J_2<0$ (FM-AFM).
\par This case is perhaps the most interesting and has attracted recent attentions~\cite{Kumar2016,PRB2024,Physica2025}. The NNN antiferromagnetic interaction plays a role of frustration and may induce instability of the ferromagnetic ground state of the nearest-neighbor ferromagnetic chain. According to the Lieb-Mattis theorem~\cite{Lieb}, the highest eigenstate of $H^{(\mathrm{isp})}$ has angular momentum $l=0$. It is well known that the ground state of $H^{(\mathrm{isp})}$ of arbitrary $N$ sites is the ferromagnetic state only if $J_1/|J_2|\geq 4$~\cite{Niemeijer1970,Ono1972,Klein1976,Bader1979,Hamada1988,Tonegawa1989,Krivnov1996,Dmitriev2006,Meisner2006,Kecke}. For $J_1/|J_2|> 4$, the ground states of $H^{(\mathrm{isp})}$ are $(N+1)$-fold degenerate and are just given by the $(N+1)$ ZESs mentioned above. Interestingly, Hamada, Kane,
Nakagawa, and Natsume found at the point $J_1/|J_2|=4$ that the ground states are actually $(N+2)$-fold degenerate: Besides the $N+1$ ZESs belonging to total angular momentum $l=N/2$, there is an additional (unnormalized) ground state with $l=0$ that can be written down explicitly,
\begin{eqnarray}\label{HKNN_GS}
|\psi_{\mathrm{HKNN}}\rangle&=&\sum_{\substack{a_j<b_j~\mathrm{and}\\a_1<a_2<\cdots<a_{N/2}}} [a_1,b_1]\cdots[a_{N/2},b_{N/2}],
\end{eqnarray}
where the sum is over all partitions of $\{1,2,\ldots,N\}$ into pairs without regard to order. Recall that the same set of partitions also appear in the definition of the Pfaffian of a skew-symmetric matrix, so there are totally $(N-1)!!=N!/\left[2^{N/2}(\frac{N}{2})!\right]$ dimer states, and hence $N!/\left(\frac{N}{2}\right)!$ Ising configurations on the right-hand side of $|\psi_{\mathrm{HKNN}}\rangle$. However, certain Ising configurations may appear multiple times, making it difficult to evaluate the norm of $|\psi_{\mathrm{HKNN}}\rangle$. It is obvious that $|\psi_{\mathrm{HKNN}}\rangle$ lies in the zero-magnetization sector. In addition, from the relation $T[i,N]=[i+1,1]=-[1,i+1]$ we have $T|\psi_{\mathrm{HKNN}}\rangle=-|\psi_{\mathrm{HKNN}}\rangle$, which means that $|\psi_{\mathrm{HKNN}}\rangle$ must carry momentum $k=-\pi$.
\par Below we focus on the case of $J_2<0$, while $J_1$ can be either positive or negative. Although the real-space wave function of $|\psi^{(\mathrm{1})}_{\mathrm{MG}}\rangle$, $|\psi^{(\mathrm{2})}_{\mathrm{MG}}\rangle$, and $|\psi_{\mathrm{HKNN}}\rangle$ seem complicated, we see that the ZESs under condition (\ref{cond}), the two degenerate MG ground states $|\psi^{(\mathrm{e})}_{\mathrm{MG}}\rangle$ and $|\psi^{(\mathrm{o})}_{\mathrm{MG}}\rangle$, as well as the HKNN ground state $|\psi_{\mathrm{HKNN}}\rangle$ all have definite momentum. It is thus interesting and desirable to reveal the momentum-space wave functions of these states. Though this is difficult for large systems, we can achieve this goal for smaller rings. Below we will take small $J_1$-$J_2$ rings of $N=6$ and $8$ sites as examples to concretely illustrate the momentum-space representations of the abovementioned particular states. 
\section{Anisotropic six-site $J_1$-$J_2$ ring}\label{SecIII}
\subsection{Block diagonalization of $H(6)$ and $\vec{L}^2$}
\par As mentioned in Sec.~\ref{SecII1}, for $N=6$ we need only to focus on magnetization sectors with $n=0,1,2$, and $3$, which allows us to exactly block diagonalize $H(6)$ by using a set of recently constructed exact two- and three-magnon Bloch states~\cite{PRB2022}. Following Ref.~\cite{PRB2022}, we define two sets of wave numbers $K_2=\left\{-2\pi/3,0,2\pi/3\right\}$ and $K'_2=\left\{-\pi,-\pi/3,\pi/3\right\}$ in the two-magnon sector. The three two-magnon Bloch states for $k\in K_2$ are constructed as~\cite{PRB2022}
\begin{eqnarray}
|\xi_1(k)\rangle&=&\frac{e^{i\frac{k}{2}}}{\sqrt{6}}\sum^5_{j=0}e^{ikj}T^j|1,2\rangle,\nonumber\\
|\xi_2(k)\rangle&=&\frac{e^{ik}}{\sqrt{6}}\sum^5_{j=0}e^{ikj}T^j|1,3\rangle,\nonumber\\
|\xi_3(k)\rangle&=&\frac{e^{i\frac{3k}{2}}}{\sqrt{3}}\sum^2_{j=0}e^{ikj}T^j|1,4\rangle,
\end{eqnarray}
where the index $r$ in $|\xi_r(k)\rangle$ measures the distance between the two down spins. The Bloch state $|\xi_3(k)\rangle$ is not defined for $k\in K'_2$.
\par Similarly, by defining $K_3=\left\{-\pi,0\right\}$ and $K'_3=\left\{-2\pi/3,-\pi/3,\pi/3,2\pi/3\right\}$ in the three-magnon sector, we can construct the following four three-magnon Bloch states for each $k\in K_3$~\cite{PRB2022}
\begin{eqnarray}
|\xi_{1,1}(k)\rangle&=&\frac{e^{ik}}{\sqrt{6}}\sum^5_{j=0}e^{ikj}T^j|1,2,3\rangle,\nonumber\\
|\xi_{1,2}(k)\rangle&=&\frac{e^{i\frac{4}{3}k}}{\sqrt{6}}\sum^5_{j=0}e^{ikj}T^j|1,2,4\rangle,,\nonumber\\
|\xi_{1,3}(k)\rangle&=&\frac{e^{i\frac{5}{3}k}}{\sqrt{6}}\sum^5_{j=0}e^{ikj}T^j|1,2,5\rangle,\nonumber\\
|\xi_{2,2}(k)\rangle&=&\frac{e^{i2k}}{\sqrt{2}}\sum^1_{j=0}e^{ikj}T^j|1,3,5\rangle,
\end{eqnarray}
where the two indices $i$ and $j$ in $|\xi_{i,j}(k)\rangle$ indicate the separations between two neighboring down spins. The Bloch state $|\xi_{2,2}(k)\rangle$ is not defined for  $k\in K'_3$.
\par The Bloch states with fixed $k$ realize a further block diagonalization of $H(6)$ in individual magnetization sectors. To obtain the corresponding Bloch Hamiltonians in the basis spanned by the Bloch states, we use the rules summarized in Appendix~\ref{AppA}~\cite{PRB2022}. The calculations are rather standard and we only present the final results. The two-magnon Bloch Hamiltonians read
\begin{eqnarray}\label{h2}
H_2(k\in K_2)&=&\left(
                \begin{array}{ccc}
                  \varepsilon_1 & t_1 & \sqrt{2}t_2 \\
                  t_1 & \varepsilon_2^- & \sqrt{2}t_1 \\
                  \sqrt{2}t_2 & \sqrt{2}t_1 & J_+ \\
                \end{array}
              \right),\nonumber\\
H_2(k\in K'_2)&=&\left(
                \begin{array}{ccc}
                  \varepsilon_1 & t_1  \\
                  t_1 & \varepsilon_2^+  \\
                \end{array}
              \right),
\end{eqnarray}
where
\begin{eqnarray}\label{onsitehopping1}
\varepsilon_1&=&-J_--J_2\cos k,~\varepsilon^\pm_2=J_-\pm J_2\cos k,\nonumber\\
t_1&=&-J_1\cos \frac{k}{2},~t_2=-J_2\cos k,
\end{eqnarray}
and the three-magnon Bloch Hamiltonians read
\begin{eqnarray}\label{h3}
H_3(k\in K_3)&=&\left(
                \begin{array}{cccc}
                  -J_- & w^*_1 & w_1 & 0 \\
                  w_1 & J_+ & w^*_2 & w_3 \\
                  w^*_1 & w_2 & J_+ & w^*_3 \\
                  0 &  w^*_3 & w_3& 3J_- \\
                \end{array}
              \right),\nonumber\\
H_3(k\in K'_3)&=&\left(
                \begin{array}{cccc}
                  -J_- & w'^*_1 & w'^*_3  \\
                  w'_1 & J_+ & w'^*_2   \\
                  w'_3 & w'_2 & J_+   \\
                \end{array}
              \right),
\end{eqnarray}
where
\begin{eqnarray}\label{onsitehopping2}
w_1&=&-\frac{J_1z}{2}-J_2z^{-2},~w_2=-J_1z-J_2z^{-2},~w_3=-\frac{\sqrt{3}}{2}J_1z,\nonumber\\
w'_1&=&-z\left(J_1/2+J_2\cos k\right),~w'_2=-(J_1z^4+J_2z^{-5})\cos k,\nonumber\\
w'_3&=&-z^5\left(J_1/2+J_2\cos k\right),
\end{eqnarray}
with $z\equiv e^{-ik/3}$.
\par  The matrix representations of the total angular momentum operator $\vec{L}^2$ in the Bloch basis can also be obtained by using the rules in Appendix~\ref{AppA}. The explicit expressions of these matrices for fixed $n$ and $k$ are given in Appendix~\ref{AppB}. By solving the eigenvalue problem of the matrices given by Eqs.~(\ref{L2matrix0})-(\ref{L2matrix3}), we obtain the possible values of $l$ in each $n$ and $k$ sector, as shown in Table I (sectors with $k$ and $-k$ share the same set of $l$'s). 
\begin{table}
\begin{center}
\begin{tabular}{ |c| c |c|c |c | }
  \hline
  & $-\pi$ & $-2\pi/3$ & $-\pi/3$ & 0 \\
  \hline
 $n=0$ &     &    &   &~~~~~~~3\\
 \hline
$n=1$  & ~~~~~~~2   &  ~~~~~2 & ~~~~~2 & ~~~~~~~3\\
\hline
 $n=2$ & ~~~~~1,2     & 1,1,2 & ~~ 1,2  &~~1,1,3\\
  \hline
 $n=3$ &  0,0,1,2  &  1,1,2 & 0,1,2 & 0,1,1,3\\
  \hline
\end{tabular}
\caption{Possible values of $l$ in each sector with fixed $n$ and $k$. Note that $l=3$ ($l=0$) lies only in the $k=0$ ($n=3$) sector.}
\end{center}
\end{table}
\par Since the lowering operator $\mathcal{L}_0$ preserves both the momentum and the angular momentum, the states with $l=3$ but different $n$'s in the last column of Table I are connected by $\mathcal{L}_0$ and belong to a unique $l=3$, the states with $k=-\pi/3$ and $l=2$ but different $n$'s belong to one of the $(l=2)$'s, etc. By counting all of these $l$'s and those in the positive $k$ sectors, we see that $l=3,2,1$, and $0$ respectively appear $1$, $5$, $9$, and $5$ times, which is consistent with Eq.~(\ref{dNl}).
\begin{table*}
\begin{center}
\begin{tabular}{ |c| c |c|c |c | }
  \hline
  & $-\pi$ & $-2\pi/3$ & $-\pi/3$ & 0 \\
  \hline
 & $-J_-+J_2$   &  0 & $-B_1/2$ & $-3(J_1+J_2)/2$\\
Eigenenergies  & $J_--J_2$   &  $(J_1+3J_2-B_2)/4$ & $B_1/2$ & $J_1-B_0/2$\\
 &        &  $(J_1+3J_2+B_2)/4$ &    & $J_1+B_0/2$\\
  \hline
 & $|\xi_1(-\pi)\rangle$  &  $ \left(
                                \begin{array}{ccc}
                                  -1 & 1 & \sqrt{2} \\
                                \end{array}
                              \right)^{\mathrm{T}}/2 $ & $|\psi^{(1)}_2(-\pi/3)\rangle$ [Eq.~(\ref{psi21pi3})] &$\left(
                                \begin{array}{ccc}
                                  \sqrt{2} & \sqrt{2} & 1 \\
                                \end{array}
                              \right)^{\mathrm{T}}/\sqrt{5}$\\
Eigenstates  & $|\xi_2(-\pi)\rangle$   &  $|\psi^{(1)}_{2,1}(-2\pi/3)\rangle$ [Eq.~(\ref{psi22pi3})] & $|\psi^{(2)}_2(-\pi/3)\rangle$ [Eq.~(\ref{psi21pi3})]  & $|\psi^{(1)}_{2,1}(0)\rangle$ [Eq.~(\ref{psi20})]\\
 &        &  $|\psi^{(2)}_{2,1}(-2\pi/3)\rangle$ [Eq.~(\ref{psi22pi3})] &    & $|\psi^{(1)}_{2,1}(0)\rangle$ [Eq.~(\ref{psi20})]\\
  \hline
   & 1  &  2 &   & 3 \\
 $l$  & 2   &  1   &     & 1\\
 &        &  1 &    & 1\\
  \hline
Notes: & type-HL & $H^{(\mathrm{isp})}$ & no definite $l$ & $H^{(\mathrm{isp})}$ \\
  \hline
\end{tabular}
\caption{Eigenenergies and eigenstates of $H(6)$ in the two-magnon sector. The values of $l$ are given for states having definite total angular momentum.}
\end{center}
\end{table*}
\subsection{Solving the Bloch Hamiltonians}
\par We are now ready to solve the Bloch Hamiltonians given by Eqs.~(\ref{h2}) and (\ref{h3}) either numerically or analytically by using for example Mathematica. As mentioned early, we focus only on the nonpositive wave numbers $k=-\pi,-2\pi/3,-\pi/3$, and $0$. There are at most $1$, $4$, $2+3+2+3=10$, and $4+3+3+4=14$ distinct eigenenergies in the $n=0,1,2$, and $3$ magnetization sectors, respectively. In turn, the full energy spectrum of $H(6)$ consists of at most $29$ different energy levels. The number of distinct eigenenergies of $H^{(\mathrm{isp})}(6)$ is reduced to $14$ since the states with the same $k$ and $l$ but different $n$'s are connected by $\mathcal{L}_0$ and are degenerate (see the last row of Table I).
\par The solutions of all the Bloch Hamiltonians $H_2(k)$ and $H_3(k)$ are detailed in Appendix~\ref{App23}. The eigenenergies and eigenstates of $H$ or $H^{(\mathrm{isp})}$ in the two-magnon sector are summarized in Table II. For $k=-\pi$, the Bloch states $|\xi_1(-\pi)\rangle$ and $|\xi_2(-\pi)\rangle$ are themselves simultaneous eigenstates of $H(6)$ and $\vec{L}^2$ and hence are type-HL states. The two eigenstates states in the $k=-\pi/3$ sector both depends on $\Delta_1$ and $\Delta_2$ and do not possess definite total angular momentum. The $3\times 3$ matrices in the $k=-2\pi/3$ and $k=0$ sectors do not admit analytical solutions unless in the isotropic case. Note that there exists a $(J_1,J_2)$-independent eigenstate of $H^{(\mathrm{isp})}(6)$ in the $k=-2\pi/3$ ($k=0$) sector since it is degenerate with the one-magnon state $|\xi(-2\pi/3)\rangle$ [$|\xi(0)\rangle$]. 
\begin{table*}
\begin{center}
\begin{tabular}{ |c| c |c|c |c | }
  \hline
  & $-\pi$ & $-2\pi/3$ & $-\pi/3$ & 0 \\
  \hline
 &  &  $J_+-(J_1+J_2)/2$ & & \\
Eigenenergies  &  $J_+-J_1+J_2$    &  $[(1+2\Delta_2)J_2+J_1-A_2]/4$ &   $J_++(J_1-J_2)/2$ & $J_++J_1+J_2$\\
 &        &  $[(1+2\Delta_2)J_2+J_1+A_2]/4$ &    &  \\
  \hline
 &   &  $\left(
           \begin{array}{ccc}
             0 & e^{i\frac{\pi}{9}} & 1 \\
           \end{array}
         \right)^{\mathrm{T}}/\sqrt{2} $ & & \\
Eigenstates  &  $\left(
                   \begin{array}{cccc}
                     0 & e^{-i\frac{\pi}{3}} & 1 & 0 \\
                   \end{array}
                 \right)^{\mathrm{T}}/\sqrt{2} $  & $|\psi^{(2)}_{3}(-2\pi/3)\rangle$ [Eq.~(\ref{psi32pi3})] & $\left(
                                                                                                                  \begin{array}{ccc}
                                                                                                                    0 & e^{i\frac{5\pi}{9}} & 1 \\
                                                                                                                  \end{array}
                                                                                                                \right)^{\mathrm{T}}/\sqrt{2}
                 $   & $\left(
                                                       \begin{array}{cccc}
                                                         0 & -1 & 1 & 0 \\
                                                       \end{array}
                                                     \right)^{\mathrm{T}}/\sqrt{2} $  \\
 &        &  $|\psi^{(3)}_{3}(-2\pi/3)\rangle$ [Eq.~(\ref{psi32pi3})] &    &  \\
  \hline
   &   &  2 &   & \\
 $l$  &  1  &  1   & 1  &  0 \\
 &        &  1 &    & \\
  \hline
\end{tabular}
\caption{The type-HL eigenstates of $H(6)$ in the three-magnon sector. }
\end{center}
\end{table*}
\par The type-HL eigenstates in the three-magnon sector are summarized in Tables III. Interestingly, all the three eigenstates in the $k=-2\pi/3$ sector are type-HL states. Actually, it can be checked that the Hamiltonian and the total angular momentum operator commute in this subspace, i.e., $[H_3(-2\pi/3),\vec{L}^2_3(-2\pi/3)]=0$, they thus have simultaneous eigenstate. The remaining three eigenstates in either the $k=-\pi$ or $k=0$ sector can only be solved in the isotropic case (see Table IV). However, the remaining two eigenstates in the $k=-\pi/3$ sector can be solved for any $\Delta_1$ and $\Delta_2$, though they do not have definite $l$.
\begin{table*}
\begin{center}
\begin{tabular}{ |c| c |c|c |c | }
  \hline
  & $-\pi$ & $-2\pi/3$ & $-\pi/3$ & 0 \\
  \hline
 & $(J_1-3J_2)/2$ &  &  $[(1+2\Delta_2)J_2-J_1-A_1]/4$ &  $-3(J_1+J_2)/2$\\
Eigenenergies  &  $J_1-A_3/2$    &    &  $[(1+2\Delta_2)J_2-J_1+A_1]/4$ & $J_1-A_0/2$\\
 &  $J_1+A_3/2$      &   &    & $J_1+A_0/2$ \\
  \hline
 &  $|\psi_{3,2}(-\pi)\rangle$ [Eq.~(\ref{psi3pi})]&   & $|\psi^{(2)}_3(-\pi/3)\rangle$ (not shown)  &  $|\psi_{3,3}(0)\rangle$\\
Eigenstates  & $|\psi^{(1)}_{3,0}(-\pi)\rangle$ [Eq.~(\ref{psi3pi})] &  & $|\psi^{(3)}_3(-\pi/3)\rangle$ (not shown) &  $|\psi^{(1)}_{3,1}(0)\rangle$   \\
 &  $|\psi^{(2)}_{3,0}(-\pi)\rangle$  [Eq.~(\ref{psi3pi})]    &   &    &  $|\psi^{(2)}_{3,1}(0)\rangle$ \\
  \hline
Notes:  &  $H^{(\mathrm{isp})}$  &      & no definite $l$  &  $H^{(\mathrm{isp})}$ \\
  \hline
\end{tabular}
\caption{The remaining eigenstates of $H(6)$ in the three-magnon sector. }
\end{center}
\end{table*}
\subsection{Energy spectrum of $H^{(\mathrm{isp})}(6)$}
\begin{figure}
\includegraphics[width=0.52\textwidth]{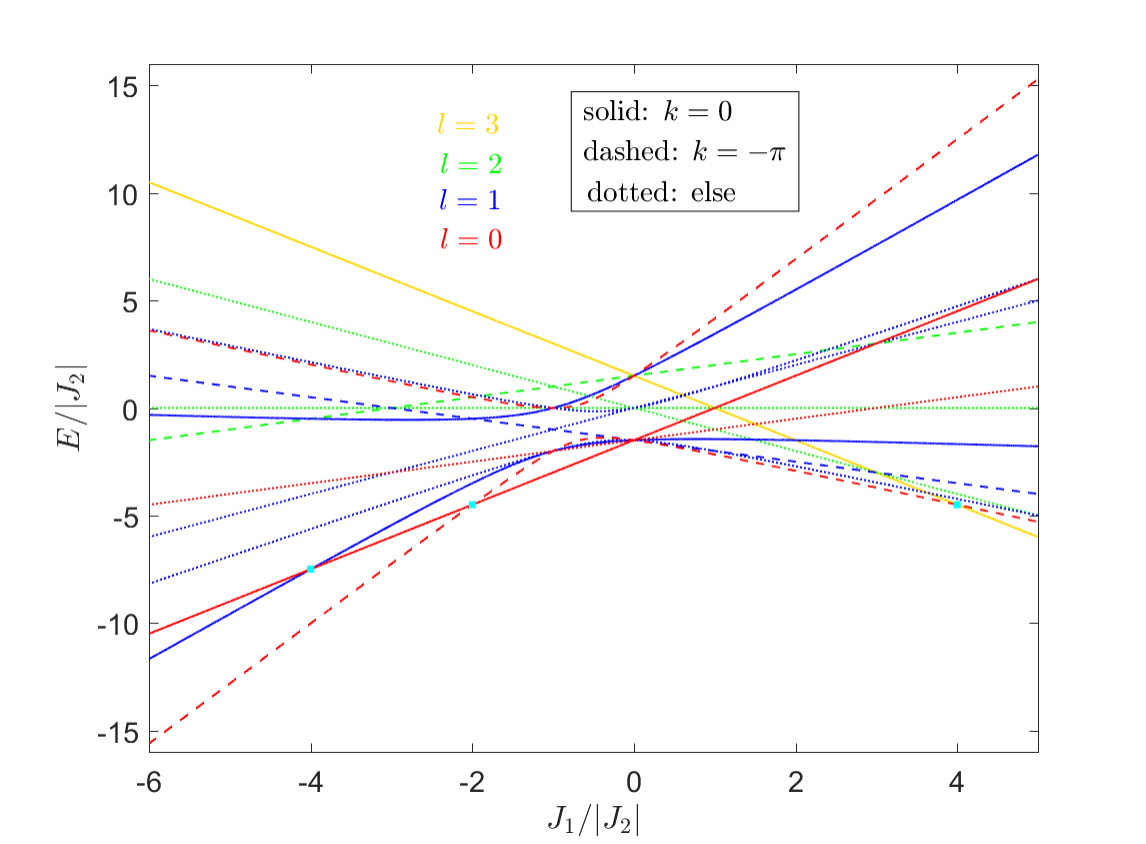}
\caption{Full spectrum of $H^{(\mathrm{isp})}(6)$ plotted by using the analytical expressions listed in Table V. Energy levels labeled by different $l$'s are plotted in different colors, while those with $k=0$ and $k=-\pi$ are respectively indicated by solid and dashed curves. There are two ground-state level crossings at the MG point and the HKNN point.}
\label{Spec14}
\end{figure}
\begin{table*}
\begin{center}
\begin{tabular}{ |c| c |c|c |c | }
  \hline
  & $-\pi$ & $-2\pi/3$ & $-\pi/3$ & 0 \\
  \hline
$n=0,~l=3$ &  &  &   &  $-3(J_1+J_2)/2$\\
  \hline
$n=1,~l=2$ & $(J_1-3J_2)/2$   & 0   & $-J_1$ &  \\
  \hline
$n=2,~l=1$ & $-(J_1-3J_2)/2$   & $(J_1+3J_2\pm B_2)/4$  & $ J_1$  & $J_1\pm B_0/2$ \\
  \hline
$n=3,~l=0$ & $J_1\pm A_3/2$   &   & $(J_1+3J_2)/2$  & $3(J_1+J_2)/2$ \\
  \hline
  \end{tabular}
\caption{Representatives of the 14 distinct eigenenergies of $H^{(\mathrm{isp})}(6)$.}
\end{center}
\end{table*}
As mentioned above, the isotropic model $H^{(\mathrm{isp})}(6)$ has at most 14 different eigenenergies. Since $ E_{0,l}(k) = E_{1,l}(k)= E_{2,l}(k)= E_{3,l}(k)$, one can choose the representatives listed in Table V for the whole spectrum. We plot these 14 eigenenergies of $H^{(\mathrm{isp})}(6)$ as functions of $J_1/|J_2|$ in Fig.~\ref{Spec14}. Different colors correspond to different total angular momenta. Two ground-state level crossings occur at the MG and HKNN point. For $J_1/|J_2|\ll0$, the ground state is a singlet carrying $k=-\pi$. As $J_1/|J_2|$ increases, a well-known triplet-singlet crossing takes place~\cite{Okamoto1992}, which is closely related to the spin fluid-dimer transition in the thermodynamic limit. This crossing can be found by setting the lowest singlet state and the lowest triplet state in the $k=0$ sector equal, i.e., $J_1-B_0/2=3(J_1+J_2)/2$, giving exactly $J_1/|J_2|=-4$. As noted by the Okamoto and Nomura, the crossing point is insensitive with respect to the change of $N$ and approaches $-4.148$ as $N\to\infty$. Note that the ground state carries $k=0$ for $-2<J_1/|J_2|<0$.   
\subsection{Momentum-space manifestation of the three types of states}
\par We now use the above analytical results to obtain explicit expressions of the ZES, the two MG ground states, and the HKNN ground state in the Bloch basis. We show that the latter two are equivalent to their real-space counterparts in terms of dimer states.
\subsubsection{Zero-energy states}
\par We take $\Delta_1=-\Delta_2=1/2$ [with $E_F=-3(J_1-J_2)/4$] as an example to illustrate the emergence of ZESs in each magnetization sector. The two wave numbers satisfying the condition are $k^*=\pm\pi/3$. It is obvious that the one-magnon eigenstate $|\psi_1(k^*)\rangle$ is a ZES since $E_1(k^*)-E_F=-(J_1-J_2)/2-J_++3J_+=0$.
\par The two ZESs in the two-magnon sector carries momentum $2k^*=\pm2\pi/3$. The corresponding Bloch Hamiltonians $H_2(\pm 2\pi/3)-E_F$ both read
\begin{eqnarray}
\left(
                \begin{array}{ccc}
                 (J_1-J_2)/2 & -J_1/2 & J_2/\sqrt{2} \\
                 -J_1/2 & J_1 & -J_1/\sqrt{2}\\
                 J_2/\sqrt{2} & -J_1/\sqrt{2} &  J_1-J_2  \\
                \end{array}
              \right).\nonumber
\end{eqnarray}
It is easy to see that the state $\left(
                                    \begin{array}{ccc}
                                      \sqrt{2} & \sqrt{2} & 1 \\
                                    \end{array}
                                  \right)^{\mathrm{T}}/\sqrt{5}
$ is an eigenstate of $H_2(\pm 2\pi/3)-E_F$ with zero eigenvalue, and hence is a ZES.
\par Since $\pi=-\pi$ in the first Brillouin zone, both of the two ZESs in the three-magnon sector carry momentum $k=-\pi$. The Bloch Hamiltonian $H_3(-\pi)-E_F$ reads
\begin{widetext}
\begin{eqnarray}
\frac{1}{2}\left(
          \begin{array}{cccc}
            J_1-2J_2  & -(J_1-2J_2)e^{-i\frac{\pi}{3}}  & -(J_1-2J_2)e^{i\frac{\pi}{3}} & 0 \\
            -(J_1-2J_2)e^{i\frac{\pi}{3}} & 2(J_1-J_2) & -2(J_1-J_2)e^{-i\frac{\pi}{3}} & -\sqrt{3}J_1e^{i\frac{\pi}{3}}  \\
            -(J_1-2J_2)e^{-i\frac{\pi}{3}}  &   -2(J_1-J_2)e^{i\frac{\pi}{3}}  & 2(J_1-J_2) & -\sqrt{3}J_1e^{-i\frac{\pi}{3}}  \\
            0 & -\sqrt{3}J_1 e^{-i\frac{\pi}{3}} &-\sqrt{3}J_1 e^{i\frac{\pi}{3}} & 3J_1 \\
          \end{array}
        \right).\nonumber
\end{eqnarray}
\end{widetext}
It can be verified that the two degenerate ZESs are given by $\left(
                                             \begin{array}{cccc}
                                               0 & 1 & e^{i\frac{\pi}{3}} & 0 \\
                                             \end{array}
                                           \right)^{\mathrm{T}}/\sqrt{2}$ and $\left(
                                             \begin{array}{cccc}
                                               2\sqrt{3} & \sqrt{3}e^{i\frac{\pi}{3}}   &\sqrt{3}e^{-i\frac{\pi}{3}}  & 2 \\
                                             \end{array}
                                           \right)^{\mathrm{T}}/\sqrt{22}$.
\subsubsection{MG ground states}
\par It can be seen from Fig.~\ref{Spec14} that at the MG point $J_2=J_1/2<0$ the two degenerate ground states are given by $|\psi^{(1)}_{3,0}(-\pi)\rangle$ with energy $E^{(1)}_{3,0}(-\pi)=J_1-A_3/2$ (see Table IV) and $|\psi_3(0)\rangle$ with energy $E_3(0)=3(J_1+J_2)/2$ (see Table III). From $A_3=-5J_1/2$ we have $E^{(1)}_{3,0}(-\pi)=E_3(0)=9J_1/4=E_{\mathrm{MG}}|_{N=6}$ and
\begin{eqnarray}
|\psi^{(1)}_{3,0}(-\pi)\rangle&=&\frac{1}{\sqrt{10}}\left(
                                                                                              \begin{array}{c}
                                                                                                0 \\
                                                                                                \sqrt{3}e^{i\frac{\pi}{3}} \\
                                                                                                \sqrt{3}e^{-i\frac{\pi}{3}} \\
                                                                                                -2\\
                                                                                              \end{array}
                                                                                            \right),\nonumber\\
                                                                                            |\psi_{3}(0)\rangle&=&\frac{1}{\sqrt{2}}\left(
                                                                                              \begin{array}{c}
                                                                                                0 \\
                                                                                                1 \\
                                                                                                -1 \\
                                                                                                0\\
                                                                                              \end{array}
                                                                                            \right).
\end{eqnarray}
\par We now show that the above two momentum-space ground states are indeed consistent with the two real-space ground states given by Eq.~(\ref{MG_GSeo}). In Appendix~\ref{AppC} we provide expressions of all the $(6-1)!!=15$ dimer states in terms of the $\binom{6}{3}=20$ real-space three-magnon basis states, which can be obtained by a direction expansion of the singlet states. Then
\begin{eqnarray}
|\psi^{(1)}_{\mathrm{MG}}\rangle&=&\left(\frac{1}{\sqrt{2}}\right)^3[12][34][56]\nonumber\\
&=&\left(\frac{1}{\sqrt{2}}\right)^3 (T^{-1}+T+T^3)(|1,2,4\rangle-|1,2,5\rangle) \nonumber\\ &&-\left(\frac{1}{\sqrt{2}}\right)^3(1-T)|1,3,5\rangle,\nonumber\\
|\psi^{(2)}_{\mathrm{MG}}\rangle&=&\left(\frac{1}{\sqrt{2}}\right)^3[23][45][61]\nonumber\\
&=&\left(\frac{1}{\sqrt{2}}\right)^3 (T^{-2}+1+T^2)(|1,2,4\rangle-|1,2,5\rangle)\nonumber\\
&& +\left(\frac{1}{\sqrt{2}}\right)^3(1-T)|1,3,5\rangle.
\end{eqnarray}
From Eq.~(\ref{MG_GSeo}) we get
\begin{eqnarray}
|\psi^{(\mathrm{e})}_{\mathrm{MG}}\rangle&=&\sqrt{\frac{2}{3}}\left(\frac{1}{\sqrt{2}}\right)^3\sum^{5}_{j=0}T^j(|1,2,4\rangle-|1,2,5\rangle)]\nonumber\\
&=&\frac{1}{\sqrt{2}}(|\xi_{1,2}(0)\rangle-|\xi_{1,3}(0)\rangle),\nonumber\\
|\psi^{(\mathrm{o})}_{\mathrm{MG}}\rangle&=& -\sqrt{\frac{2}{5}}\left(\frac{1}{\sqrt{2}}\right)^3 \sum^5_{j=0}e^{i(-\pi)j}T^j(|1,2,4\rangle-|1,2,5\rangle)] \nonumber\\
&&-\sqrt{\frac{2}{5}}\frac{1}{\sqrt{2}}\sum^1_{j=0}e^{i(-\pi)j}|1,3,5\rangle\nonumber\\
&=& \sqrt{\frac{3}{10}}[e^{i\frac{\pi}{3}}|\xi_{1,2}(-\pi)\rangle+e^{-i\frac{\pi}{3}}|\xi_{1,3}(-\pi)\rangle] \nonumber\\ &&-\sqrt{\frac{2}{5}}|\xi_{2,2}(-\pi)\rangle,
\end{eqnarray}
giving
\begin{eqnarray}
|\psi^{(\mathrm{o})}_{\mathrm{MG}}\rangle&=&|\psi^{(1)}_{3,0}(-\pi)\rangle,~|\psi^{(\mathrm{e})}_{\mathrm{MG}}\rangle=|\psi_{3}(0)\rangle.
\end{eqnarray}
\subsubsection{HKNN ground state}
\par From Fig.~\ref{Spec14} we see that the HKNN ground state at $J_1=-4J_2>0$ is also given by $|\psi^{(1)}_{3,0}(-\pi)\rangle$ with energy $E^{(1)}_{3,0}(-\pi)=-9J_1/8=E_F$~\cite{Hamada1988}. From Eq.~(\ref{psi3pi}), the corresponding ground state is
\begin{eqnarray}\label{psi30HKNN}
|\psi^{(1)}_{3,0}(-\pi)\rangle&=&\frac{1}{\sqrt{34}}\left(
                                                                                              \begin{array}{c}
                                                                                                3\sqrt{3} \\
                                                                                                \sqrt{3}e^{i\frac{\pi}{3}} \\
                                                                                                \sqrt{3}e^{-i\frac{\pi}{3}} \\
                                                                                                1\\
                                                                                              \end{array}
                                                                                            \right).
\end{eqnarray}
\par Using the expressions for the dimer states given by Eq.~(\ref{dimermagnon}) and collecting all the like terms, we have
\begin{eqnarray}
&&|\psi_{\mathrm{HKNN}}\rangle\nonumber\\
&=&[12][34][56]+[12][35][46]+[12][36][45] +[13][24][56]\nonumber\\
&&+[13][25][46]+[13][26][45]+[14][23][56] +[14][25][36] \nonumber\\
&&+[14][26][35] +[15][23][46]+[15][24][36]+[15][26][34]\nonumber\\
&&+[16][23][45]+[16][24][35]+[16][25][34]\nonumber\\
&=&-2\sum^5_{j=0}e^{i(-\pi) j}T^j(3|1,2,3\rangle+|1,2,4\rangle-|1,2,5\rangle)\nonumber\\
&& +2\sum^1_{j=0}e^{i(-\pi) j}T^j|1,3,5\rangle\nonumber\\
&=&6\sqrt{6}|\xi_{1,1}(-\pi)\rangle+2\sqrt{6}e^{i\frac{\pi}{3}}|\xi_{1,2}(-\pi)\rangle\nonumber\\
&& +2\sqrt{6} e^{-i\frac{\pi}{3}}|\xi_{1,3}(-\pi)\rangle+2\sqrt{2}|\xi_{2,2}(-\pi)\rangle.
\end{eqnarray}
As expected, the state $|\psi_{\mathrm{HKNN}}\rangle$ is not normalized. Indeed, from Eq.~(\ref{psi30HKNN}) we have
\begin{eqnarray}
|\psi_{\mathrm{HKNN}}\rangle=4\sqrt{17} |\psi^{(1)}_{3,0}(-\pi)\rangle.
\end{eqnarray}
The momentum-space wave function of $|\psi_{\mathrm{HKNN}}\rangle$ tells us that the Bloch state $|\xi_{1,1}(-\pi)\rangle=-\frac{1}{\sqrt{6}}(|1,2,3\rangle-|2,3,4\rangle+|3,4,5\rangle-|4,5,6\rangle+|5,6,1\rangle-|6,1,2\rangle)$ has the largest weight, which means that the HKNN ground state behaves like a ``bound state" with three successive down spins binding together. We will see later that this property also holds for $N=8$.
\section{Isotropic eight-site $J_1$-$J_2$ ring}\label{SecIV}
\subsection{Block diagonalization}
\par We now turn to study an isotropic eight-site $J_1$-$J_2$ ring with $\Delta_1=\Delta_2=1$. As mentioned above, the four-magnon subspace with $l_z=0$ contains all the eigenvalues of $H(8)$ due to the degeneracy among the eigenstates connected by $\mathcal{L}_0$. We thus focus on this zero-magnetization subspace below. Among the $\binom{8}{4}=70$ states, there are $\binom{8}{4}-\binom{8}{3}=14$ ones belonging to $l=0$ since the three-magnon sector contains all the states with $l\neq 0$. For an antiferromagnetic XXX ring with $J_1=-1$ and $J_2=0$, Hulth\'en was able to decompose the $14$-dimensional $l=0$ subspace into smaller subspaces using a certain set of dimer states~\cite{Hulthen}. This motivated Majumdar and Chosh to try to perform a similar decomposition for the eight-site $J_1$-$J_2$ ring, though they failed to do so~\cite{MG}. As we will see, by properly defining a set of four-magnon Bloch states, we can decompose the whole 70-dimensional subspace into smaller blocks of at most 4 dimensions using the two quantum numbers $l$ and $k$. 
\par Although the four-magnon Bloch state for a general $N$-site ring has not been systematically constructed, we can explicitly write them down in the case of an eight-site spin-1/2 XXZ or $J_1$-$J_2$ ring. For $N=8$, we have to define three sets of wave numbers, i.e., $K_4=\{-\pi,0\}$, $\tilde{K}_4=\{-\frac{\pi}{2},\frac{\pi}{2}\}$, and $K'_4=\{-\frac{3}{4}\pi,~-\frac{\pi}{4},\frac{\pi}{4},~\frac{3}{4}\pi\}$. 
\begin{figure}
\includegraphics[width=0.52\textwidth]{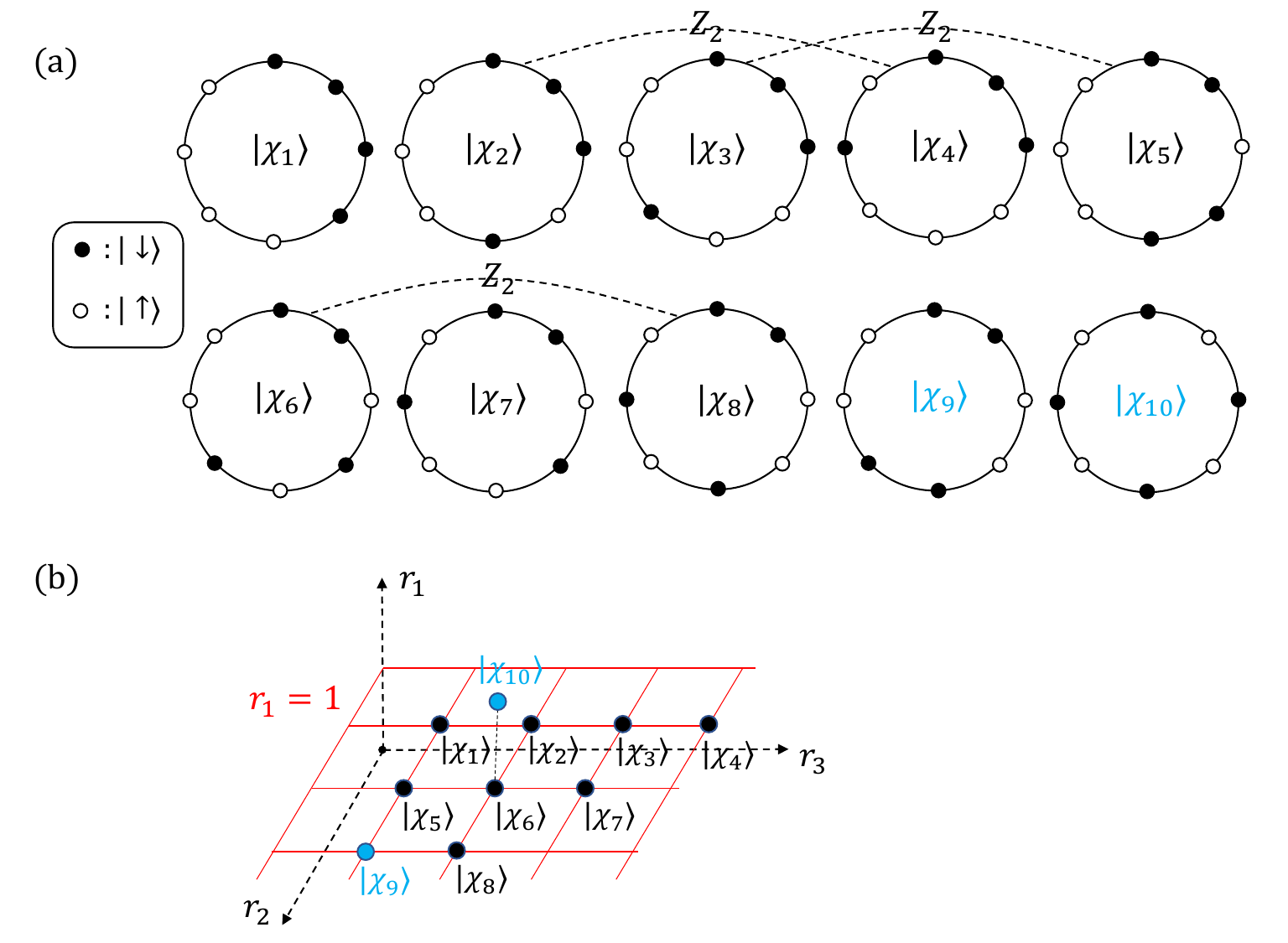}
\caption{(a) The ten local states $|1,1+r_1,1+r_1+r_2,1+r_1+r_2+r_3\rangle$ from which we construct the ten four-magnon Bloch states $|\chi_i(k)\rangle$ ($i=1,2,\cdots,10$) for $k\in K_4$. The dashed curves link two Bloch states connected by the $\mathbb{Z}_2$ transformation. (b) The ten Bloch states defines a ten-dimensional single-particle problem on the $r_1$-$r_2$-$r_3$ lattice.}
\label{4magnonBloch}
\end{figure}
\par For $k\in K_4$, we construct the following four-magnon Bloch states
\begin{eqnarray}
&&|\xi_{r_1,r_2,r_3}(k)\rangle=\frac{e^{(3r_1+2r_2+r_3)i\frac{k}{4}}}{\sqrt{N}}\sum^{N-1}_{n=0}e^{ikn}\nonumber\\
&&T^n|1,1+r_1,1+r_1+r_2,1+r_1+r_2+r_3\rangle,
\end{eqnarray}
where $(r_1,r_2,r_3)=(1,1,1)$, $(1,1,2)$, $(1,1,3)$, $(1,1,4)$, $(1,2,1$), $(1,2,2)$, $(1,2,3)$, and $(1,3,1)$. For  $(r_1,r_2,r_3)=(1,3,2)$ and $(2,2,2)$, we respectively define
\begin{eqnarray}
&&|\xi_{1,3,1}(k)\rangle=\frac{e^{i\frac{5k}{2}}}{2}\sum^{3}_{n=0}e^{ikn}T^n|1,2,5,6\rangle,
\end{eqnarray}
and
\begin{eqnarray}
&&|\xi_{2,2,1}(k)\rangle=\frac{e^{i3k}}{\sqrt{2}}\sum^{1}_{n=0}e^{ikn}T^n|1,3,5,7\rangle.
\end{eqnarray}
The ten local states $|1,1+r_1,1+r_1+r_2,1+r_1+r_2+r_3\rangle$ are shown in Fig.~\ref{4magnonBloch}(a). To simplify the notations, we define the ordered Bloch basis states $|\chi_1(k)\rangle=|\xi_{1,1,1}(k)\rangle, |\chi_2(k)\rangle=|\xi_{1,1,2}(k)\rangle,\ldots,|\chi_{9}(k)\rangle=|\xi_{1,3,1}(k)\rangle$, and $|\chi_{10}(k)\rangle=|\xi_{2,2,2}(k)\rangle$, which can be represented by the ten lattice points in the $r_1$-$r_2$-$r_3$ space [see Fig.~\ref{4magnonBloch}(b)]. 
\par From the conservation of the total angular momentum for an isotropic ring, the ten-dimensional Bloch Hamiltonian with fixed $k=0$ or $-\pi$ can be further split into several smaller blocks having fixed $l$. The total angular momentum $\vec{L}^2$ for $N=8$ reads
\begin{eqnarray}
\vec{L}^2&=&4+L^2_z+X_1+X_2+X_3+X_4,
\end{eqnarray}
where $X_i=(S^+_1S^-_{1+i}+S^+_2S^-_{2+i}+\cdots+S^+_8S^-_{i})+\mathrm{H.c.}$ if $i=1,2,3$ and $X_4=(S^+_1S^-_{5}+S^+_2S^-_{6}+S^+_3S^-_{7}+S^+_8S^-_{4})+\mathrm{H.c.}$.
\par Using the rules outlined in Appendix~\ref{AppA}, it is straightforward to obtain the $10\times10$ matrix representations of both $H^{(\mathrm{isp})}(8)$ and $\vec{L}^2$ in the ordered basis $\{|\chi_1(k)\rangle,\ldots,|\chi_{10}(k)\rangle\}$, whose explicit expressions are omitted here for simplicity. Similarly, for $k\in \tilde{K}_4$ ($k\in K'_4$) we get $9\times9$ ($8\times 8$) matrices since $|\chi_{10}(k)\rangle$ is ($|\chi_{9}(k)\rangle$ and $|\chi_{10}(k)\rangle$ are) not well defined. 
\par The matrix representations of $\vec{L}^2$ can be directly solved by Mathematica. The resulting values of $l$ in each $k$-subspace are listed in the last row of Table VI. There are totally 45 columns in Table VI, which means that the number of distinct eigenenergies is at most 45. For completeness, the $l$'s in the zero-, one-, two-, and three-magnon sectors are given in the first four rows of the table. Since $[H^{(\mathrm{isp})},\vec{L}^2]=0$, the Hamiltonian forms a smaller block in each degenerate subspace labeled by fixed $l$ and $k$~\cite{PhysEng}. 
\begin{table*}
\begin{center}
\begin{tabular}{ |c|c|c|c|c|c| }
  \hline
    & $-\pi$ & $-3\pi/4$ & $-\pi/2$ & $-\pi/4$ & 0 \\
  \hline
 $n=0$ &     &    &   &~~~~~~~  &~~~~~~~~~~~~~~~~~~~~~4\\
 \hline
$n=1$  & ~~~~~~~~~~~~~~~~~~~~~3& ~~~~~~~~~~~~~~~~3 & ~~~~~~~~~~~~~~~~~~~3 & ~~~~~~~~~~~~~~~~3 &~~~~~~~~~~~~~~~~~~~~~4\\
\hline
 $n=2$ &~~~~~~~~~~~~~~2,2,2,3& ~~~~~~~~~~~\textcolor{blue}{\textbf{2},\textbf{2}},3 &~~~~~~~~~~~~\textcolor{blue}{2,2,2},3&~~~~~~~~~~~\textcolor{blue}{\textbf{2},\textbf{2}},3 &~~~~~~~~~~~~~~\textcolor{blue}{2,2,2},4\\
  \hline
 $n=3$ &~~~~~~~1,1,1,2,2,2,3 & ~~1,1,1,1,2,2,3 &~~~~~1,1,1,2,2,2,3 & ~~1,1,1,1,2,2,3 &~~~~~~~1,1,1,2,2,2,4\\
  \hline
   $n=4$ &\textcolor{blue}{0,0,0,1,1,1,\textbf{2},\textbf{2},\textbf{2},\textbf{3}} & \textcolor{blue}{\textbf{0},1,1,1,1},2,2,\textcolor{blue}{\textbf{3}}&\textcolor{blue}{\textbf{0},\textbf{0},1,1,1},2,2,2,\textcolor{blue}{\textbf{3}}& \textcolor{blue}{\textbf{0},1,1,1,1},2,2,\textcolor{blue}{\textbf{3}}&\textcolor{blue}{0,0,0,\textbf{1},\textbf{1},\textbf{1},2,2,2,\textbf{4}}\\
  \hline
\end{tabular}
\caption{Possible values of $l$ in each sector with fixed $n$ and $k$ for an isotropic $J_1$-$J_2$ chain of $N=8$ sites. The representatives of the energy levels labelled by fixed $k$ and $l$ are indicated in blue, among which the eigenenergies expressible in terms of linear functions or square roots are shown in bold. Although there are totally 45 columns in the table, the special eigenenergy of value $J_2$ is shared by $(k,n,l)=(-\pi,4,2)$ and $(0,4,1)$, so that the number of distinct eigenenergies are at most 44.}
\end{center}
\end{table*}
\subsubsection{$k=-\pi$}
For $k=-\pi$, the orthonormal eigenstate $|\eta_l(-\pi)\rangle$ of $\vec{L}^2$ belonging to the eigenvalue $l(l+1)$ can be solved as (in the Bloch basis $\{|\chi_i(-\pi)\rangle\}$)
\begin{eqnarray}
|\eta_3\rangle&=&\frac{1}{\sqrt{5}}(0,i,0,-1,0,1,0,-i,0,e^{-i\frac{\pi}{4}})^{\mathrm{T}},
\end{eqnarray}
\begin{eqnarray}
|\eta^{(1)}_2\rangle&=&\frac{1}{\sqrt{2}}(0,0,-1,0,1,0,0,0,0,0)^{\mathrm{T}},\nonumber\\
|\eta^{(2)}_2\rangle&=&\frac{1}{\sqrt{6}}(1,-e^{i\frac{\pi}{4}},0,-e^{-i\frac{\pi}{4}},0,0,-1,0,\sqrt{2},0)^{\mathrm{T}},\nonumber\\
|\eta^{(3)}_2\rangle&=&\frac{1}{\sqrt{6}}(-e^{-i\frac{\pi}{4}},0,0,0,0,-i,e^{-i\frac{\pi}{4}},1,\sqrt{2}e^{-i\frac{\pi}{4}},0)^{\mathrm{T}},\nonumber\\
\end{eqnarray}
\begin{eqnarray}
|\eta^{(1)}_1\rangle&=&\frac{1}{\sqrt{2}}(0,0,1,0,1,0,0,0,0,0)^{\mathrm{T}},\nonumber\\
|\eta^{(2)}_1\rangle&=&\frac{1}{2}(0,1,0,i,0,i,0,1,0,0)^{\mathrm{T}},\nonumber\\
|\eta^{(3)}_1\rangle&=&\frac{1}{2\sqrt{5}}(0,e^{-i\frac{\pi}{4}},0,e^{i\frac{\pi}{4}},0,-e^{i\frac{\pi}{4}},0,-e^{-i\frac{\pi}{4}},0,4)^{\mathrm{T}},\nonumber\\
\end{eqnarray} 
and
\begin{eqnarray}
|\eta^{(1)}_0\rangle&=&\frac{1}{\sqrt{2}}(1,0,0,0,0,0,1,0,0,0)^{\mathrm{T}},\nonumber\\
|\eta^{(2)}_0\rangle&=&\frac{1}{\sqrt{6}}(1,e^{i\frac{\pi}{4}},0,e^{-i\frac{\pi}{4}},0,e^{-i\frac{\pi}{4}},-1,e^{i\frac{\pi}{4}},0,0)^{\mathrm{T}},\nonumber\\
|\eta^{(3)}_0\rangle&=&\frac{1}{\sqrt{6}}(0,e^{i\frac{\pi}{4}},0,e^{-i\frac{\pi}{4}},0,-e^{-i\frac{\pi}{4}},0,-e^{i\frac{\pi}{4}},\sqrt{2},0)^{\mathrm{T}}.\nonumber\\
\end{eqnarray} 
The corresponding block Hamiltonians read
\begin{eqnarray}
h_3(-\pi)=-2J_2
\end{eqnarray}
\begin{eqnarray}
h_2(-\pi)=\left(
           \begin{array}{ccc}
             J_2 & 0 & 0 \\
             0 & J_2 & J_2e^{-i\frac{\pi}{4}} \\
             0 & J_2e^{i\frac{\pi}{4}} & J_2-J_1 \\
           \end{array}
         \right),
\end{eqnarray}
\begin{eqnarray}
h_1(-\pi)=\left(
           \begin{array}{ccc}
             J_2 & -\sqrt{2}J_1e^{i\frac{\pi}{4}} & 0 \\
             -\sqrt{2}J_1e^{-i\frac{\pi}{4}} &\frac{J_1}{2}& -\frac{\sqrt{5}}{2}e^{-i\frac{\pi}{4}}\\
             0 & -\frac{\sqrt{5}}{2}e^{i\frac{\pi}{4}} & \frac{5}{2}J_1+-2J_2 \\
           \end{array}
         \right),
\end{eqnarray} 
and
\begin{eqnarray}\label{h0_pi}
h_0(-\pi)=\left(
           \begin{array}{ccc}
             0 & -\sqrt{3}J_1 & 0 \\
             -\sqrt{3}J_1 & J_1 & -J_1 \\
             0 & -J_1 & J_1+2J_2 \\
           \end{array}
         \right).
\end{eqnarray} 
We note that $h_2(-\pi)$ can be solved analytically, giving three eigenenergies 
\begin{eqnarray}
E^{(1)}_2(-\pi)&=&J_2,\nonumber\\
E^{(2)}_2(-\pi)&=&J_2-\frac{J_1}{2}-\frac{1}{2}\sqrt{J^2_1+4J^2_2},\nonumber\\
E^{(3)}_2(-\pi)&=&J_2-\frac{J_1}{2}+\frac{1}{2}\sqrt{J^2_1+4J^2_2}.
\end{eqnarray} 
However, in general $h_1(-\pi)$ and $h_0(-\pi)$ have no analytical solutions except for several particular cases, which will be discussed in detail below.
\subsubsection{$k=0$}
\par Similar calculations give the following block Hamiltonians in the $k=0$ subspace:
\begin{eqnarray}
h_4(0)=-2J_1-2J_2=E_F,
\end{eqnarray}
\begin{eqnarray}
h_2(0)=\left(
           \begin{array}{ccc}
             \frac{3}{7}(4J_1-3J_2) & \frac{1}{7\sqrt{5}}(J_1+8J_2) & \frac{1}{\sqrt{5}}(J_1-2J_2) \\
             \frac{1}{7\sqrt{5}}(J_1+8J_2) & \frac{1}{35}(31J_2-4J_1) & \frac{1}{5}(J_2-4J_1) \\
             \frac{1}{\sqrt{5}}(J_1-2J_2) & \frac{1}{5}(J_2-4J_1)  & -\frac{3}{5}(J_1+J_2) \\
           \end{array}
         \right),
\end{eqnarray}
\begin{eqnarray}
h_1(0)=\left(
           \begin{array}{ccc}
             J1+J_2 & 0 & -J_2 \\
             0 & J_2 & 0\\
             -J_2 & 0 & J_2 \\
           \end{array}
         \right),
\end{eqnarray} 
and
\begin{eqnarray}\label{h0_0}
h_0(0)=\left(
           \begin{array}{ccc}
             J_1 & -J_1/\sqrt{3} & -\sqrt{\frac{5}{3}}J_1 \\
             -J_1/\sqrt{3} & 4J_2 & 0\\
             -\sqrt{\frac{5}{3}}J_1 & 0 & 3J_1-2J_2 \\
           \end{array}
         \right).
\end{eqnarray} 
Solving $h_1(0)$ gives three eigenvalues 
\begin{eqnarray}
E^{(1)}_1(0)&=&J_2,\nonumber\\
E^{(2)}_1(0)&=&J_2+\frac{J_1}{2}-\frac{1}{2}\sqrt{J^2_1+4J^2_2},\nonumber\\
E^{(3)}_1(0)&=&J_2+\frac{J_1}{2}+\frac{1}{2}\sqrt{J^2_1+4J^2_2}.
\end{eqnarray} 
Again, $h_0(0)$ is soluble at special values of $J_1/|J_2|$. Here, we note that $E^{(1)}_2(-\pi)=E^{(1)}_1(0)=J_2$, which means that there are actually at most 44 distinct eigenenergies. 
\subsubsection{$k=-\pi/2$}
\par The three eigenenergies belonging to $l=2$ can be more easily obtained in the two-magnon sector (see the third row of Table VI). We only present the block Hamiltonians for $l=3,1$, and $0$ in the four-magnon sector:
\begin{eqnarray}
h_3(-\pi/2)=-J_1,
\end{eqnarray}
\begin{eqnarray}
&&h_1(-\pi/2)\nonumber\\
&=&\left(
           \begin{array}{ccc}
             \frac{2J_1+J_2}{3} & -\frac{\sqrt{2}e^{-i\frac{\pi}{8}}(2J_1+J_2)}{6} & -\sqrt{\frac{5}{6}}J_2 \\
             -\frac{\sqrt{2}e^{i\frac{\pi}{8}}(2J_1+J_2)}{6} & \frac{13J_1+2J_2}{12} & \sqrt{\frac{5}{3}}\frac{e^{\frac{i\pi}{8}}(2J_2- 3J_1)}{4}\\
             -\sqrt{\frac{5}{6}}J_2 & \sqrt{\frac{5}{3}}\frac{e^{-\frac{i\pi}{8}}( 2J_2-3J_1)}{4} & \frac{J_1+10J_2}{4}  \\
           \end{array}
         \right),
\end{eqnarray} 
\begin{eqnarray}
h_0(-\pi/2)=\left(
                  \begin{array}{cc}
                    2J_2 & e^{i\frac{3\pi}{8}}J_1 \\
                    e^{-i\frac{3\pi}{8}}J_1 & J_1 \\
                  \end{array}
                \right).
\end{eqnarray} 
It is easy to diagonalize $h_0(-\pi/2)$ to get
\begin{eqnarray}
E^{(1)}_0(-\pi/2)&=&\frac{J_1}{2}+J_2+\frac{1}{2}\sqrt{5J^2_1-4J_1J_2+J^2_2},\nonumber\\
E^{(2)}_0(-\pi/2)&=&\frac{J_1}{2}+J_2-\frac{1}{2}\sqrt{5J^2_1-4J_1J_2+J^2_2}.
\end{eqnarray} 
It should be mentioned that Majumdar and Chosh also obtained the above two eigenenergies, though the corresponding wave number $k$ was not identified there~\cite{MG}.
\subsubsection{$k=-3\pi/4$ and $-\pi/4$}
\par The eigenenergies and eigenstates in these two subspaces share similar structures. The eigenenergies belonging to $l=2$ are given by 
\begin{eqnarray}
E^{(1)}_2(-3\pi/4)&=&-\frac{\sqrt{2}}{4}(J_1+\sqrt{5J^2_1-8J_1J_2+8J^2_2}),\nonumber\\
E^{(2)}_2(-3\pi/4)&=&-\frac{\sqrt{2}}{4}(J_1-\sqrt{5J^2_1-8J_1J_2+8J^2_2}),
\end{eqnarray} 
and 
\begin{eqnarray}
E^{(1)}_2(-\pi/4)&=&\frac{\sqrt{2}}{4}(J_1+\sqrt{5J^2_1-8J_1J_2+8J^2_2}),\nonumber\\
E^{(2)}_2(-\pi/4)&=&\frac{\sqrt{2}}{4}(J_1-\sqrt{5J^2_1-8J_1J_2+8J^2_2}).
\end{eqnarray} 
The eigenenergies belonging to $l=0$ are given by 
\begin{eqnarray}
E_0(-3\pi/4)&=&\frac{2+\sqrt{2}}{2}J_1+J_2,\nonumber\\
E_0(-\pi/4)&=&\frac{2-\sqrt{2}}{2}J_1+J_2.
\end{eqnarray} 
These two eigenenergies are also found by Majumdar and Chosh~\cite{MG}. 
\par It is difficult to write down analytical expressions for the four-dimensional block Hamiltonians in the $l=1$ subspace, so we numerically calculate the corresponding eigengeneries $E^{(i)}_1(-3\pi/4)$ and $E^{(i)}_1(-\pi/4)$ ($i=1,2,3,4$).
\begin{figure}
\includegraphics[width=0.52\textwidth]{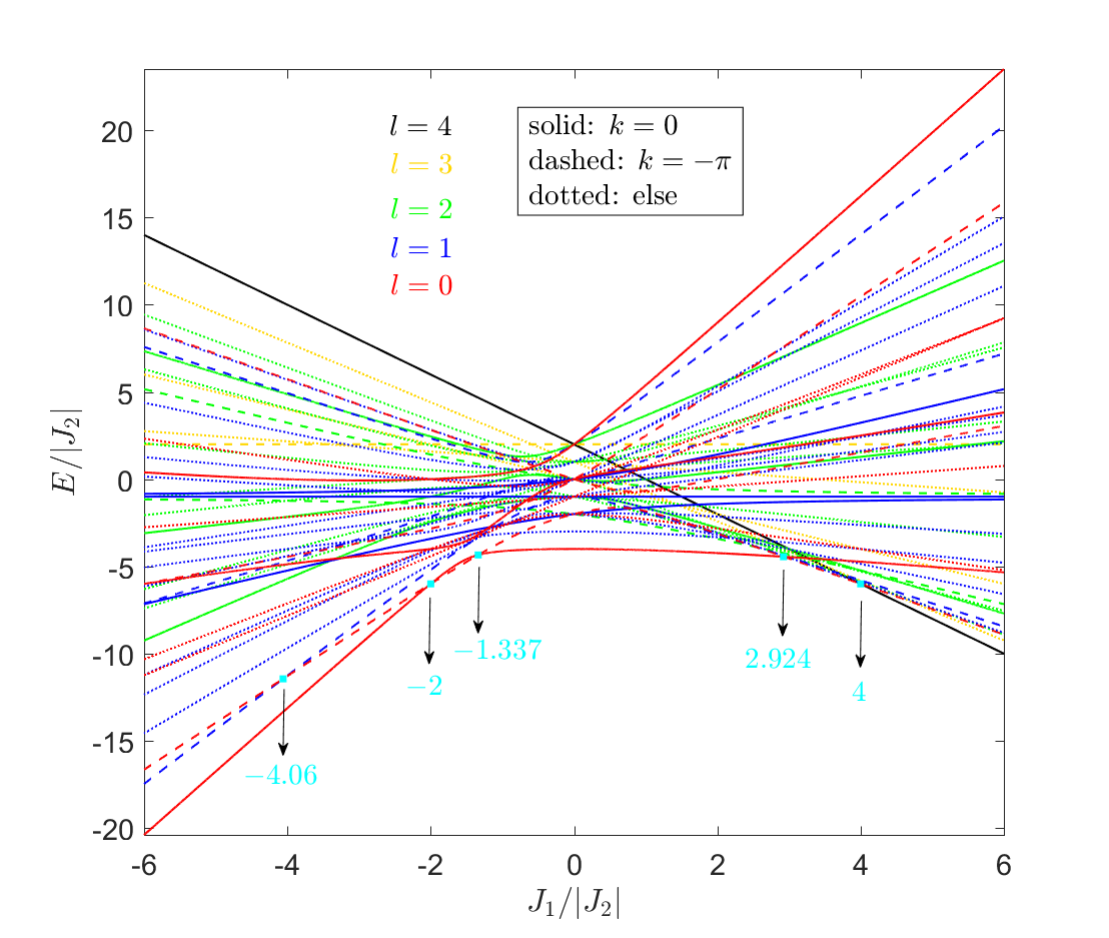}
\caption{Full spectrum of $H^{(\mathrm{isp})}(8)$ plotted by using the block Hamiltonians with fixed $l$'s. Energy levels labeled by different $l$'s are plotted in different colors, while those with $k=0$ and $k=-\pi$ are respectively indicated by solid and dashed curves. There are four ground-state level crossings at $J_1/|J_2|=-2,-1.337,~2.924$, and $4$. A singlet-triplet crossing in the first excited state also occurs at $J_1/|J_2|=-4.06$.}
\label{Spec8}
\end{figure}
\subsection{Full spectrum, ground-state wave function, and spin correlations}
\par The full spectrum of $H^{(\mathrm{isp})}(8)$ calculated by diagonalizing the above block Hamiltonians are shown in Fig.~\ref{Spec8}, where energy levels belonging to different $l$'s ($k$'s) are indicated in different colors (line styles). Due to the degeneracy of eigenstates belonging to different magnetization sectors, the spectrum is completely specified by the two quantum numbers $k$ and $l$. The ground state has angular momentum $l=0$ ($l=4$) for $J_1/|J_2|<4$ ($J_1/|J_2|>4$). For $J_1/|J_2|\ll 0$, the system is more like an antiferromagnetic XXX ring where the ground state is a singlet carrying $k=0$ and the first excited state is a triplet carrying $k=-\pi$. As $J_1/|J_2|$ increases, the triplet-singlet crossing occurs at $J_1/|J_2|=-4.06$~\cite{Okamoto1992}. Similar to the case of $N=6$, a change of $k$ occurs at the MG point $J_1/|J_2|=-2$, which is followed by the transition back to $k=0$ at $J_1/|J_2|=-1.337$~\cite{TonegawaAFM}. When $J_1/|J_2|$ is increased to $2.924$, the wave number turns to $k=-\pi$ again until reaching the HKNN point. 
\par
\begin{figure}
\includegraphics[width=0.52\textwidth]{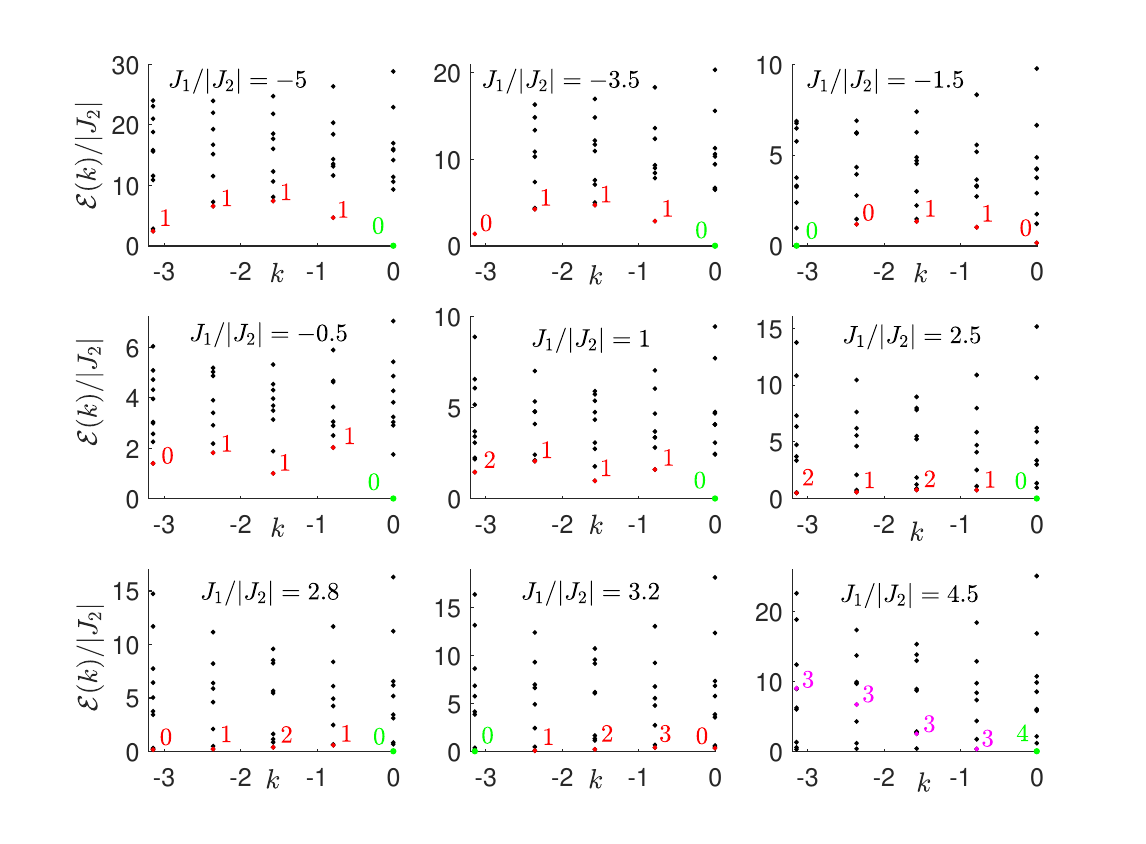}
\caption{The full excitation spectrum $\mathcal{E}(k)/|J_2|=[E(k)-E_{\mathrm{GS}}]/|J_2|$ for various values of $J_1/|J_2|$. The lowest excitation energy for each $k$ is indicated in red and the green dot corresponds to the ground state. The quantum number $l$ is shown beside each of the above dots. In the last panel with $J_1/|J_2|=4.5$, the ground state is ferromagnetic and the spin-wave spectrum with $l=3$ is indicated in pink.}
\label{spinwave}
\end{figure}

 For fixed $J_1/|J_2|$, the full excitation spectrum $\mathcal{E}(k)/|J_2|=[E(k)-E_{\mathrm{GS}}]/|J_2|$ (with $E_{\mathrm{GS}}$ the ground-state energy) as a function of $k$ is presented in Fig.~\ref{spinwave}. For $J_1/|J_2|=-5$, the lowest excitation energy $\mathcal{E}_{\min}(k)/|J_2|$ (red dots) behaves roughly like the Cloizeaux-Pearson dispersion $\sim |\sin k|$, which arises from two-spinon states in the case of an antiferromagnetic XXX ring~\cite{Cloizeaux1962,FT81}, which carries $l=1$ as expected. For $J_1/|J_2|=-3.5$, the ground state lies in the dimer phase and the lowest-energy state in the $k=-\pi$ subspace becomes a singlet. As $J_1/|J_2|$ increases, the profile of $\mathcal{E}_{\min}(k)/|J_2|$ starts to show multiple maxima (see the middle panels of Fig.~\ref{spinwave}), accompanied by the change of the corresponding $l$ to higher values such as $l=2$ and $l=3$. For $J_1/|J_2|=4.5$ (the last panel), the ground state becomes ferromagnetic and the ``spin-wave" states with $l=3$~\cite{Mattheiss1961} are indicated in pink. 
\par
\begin{figure}
\includegraphics[width=0.52\textwidth]{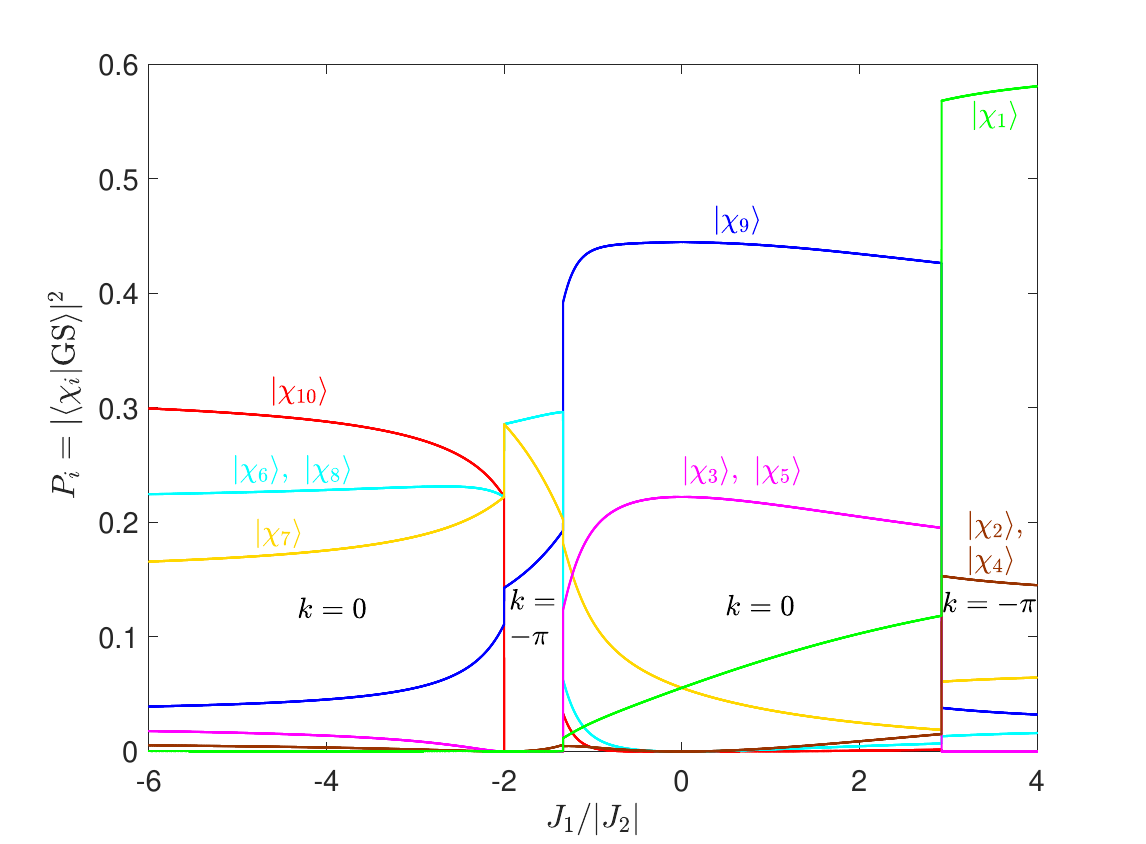}
\caption{The weight $P_i=|\langle\chi_i(k)|\mathrm{GS}\rangle|^2$ of the Bloch state $|\chi_i(k)\rangle$ in the ground state $|\chi_i(k)\rangle$.}
\label{weight}
\end{figure}

 We next discuss the ground-state wave function in the momentum space. Figure \ref{weight} shows the weight of the Bloch state $|\chi_i(k)\rangle$ in the ground state $|\mathrm{GS}\rangle$, i.e., $P_i=|\langle\chi_i(k)|\mathrm{GS}\rangle|^2$. Note that $P_i$ is discontinuous at the level crossings $J_1/|J_2|=-2,-1.337$, and $2.924$. Two Bloch states connected by the $\mathbb{Z}_2$ transformation (up to a phase factor), such as $|\chi_2(k)\rangle$ and $|\chi_4(k)\rangle$, have the same weight. The highest weighted state in each region, i.e., $|\chi_{10}(k)\rangle$, $|\chi_{6}(k)\rangle$ and $|\chi_{8}(k)\rangle$, $|\chi_{9}(k)\rangle$, and $|\chi_{1}(k)\rangle$, roughly reflects the nature the ground state. However, $N=8$ is too small to find the hint of the fluid-dimer transition from the weights. Note that at the MG point the weights of some Bloch states become equal while some others vanishes exactly. Also, at the HKNN point the weights of $|\chi_{3}(-\pi)\rangle$, $|\chi_{5}(-\pi)\rangle$, and $|\chi_{10}(-\pi)\rangle$ are exactly zero. These facts indicates that the MG ground states and HKNN ground state might admit analytical forms. We will see later that this is the case. For large enough rings, Soos, Parvej, and Kumar~\cite{Kumar2016} have shown that the middle region $-2\leq J_1/|J_2|\leq 4$ can be divided into three intervals: a decoupled commensurate phase in $[-0.44,1.24]$ and two incommensurate gapped phases in $[-2,-0.44]$ and $[1.24,4]$. Again, $N=8$ is so small that we cannot see any feature of these phases based on the weights and level crossings. An interesting fact to note is that the Bloch state $|\chi_{1}(-\pi)\rangle=|\xi_{1,1,1}(-\pi)\rangle$ is largely weighted at the HKNN point, which means that the HKNN state may also behave like a four-magnon bound state. We will see this point later from the analytical expression of the HKNN ground state.
\par
\begin{figure}
\includegraphics[width=0.52\textwidth]{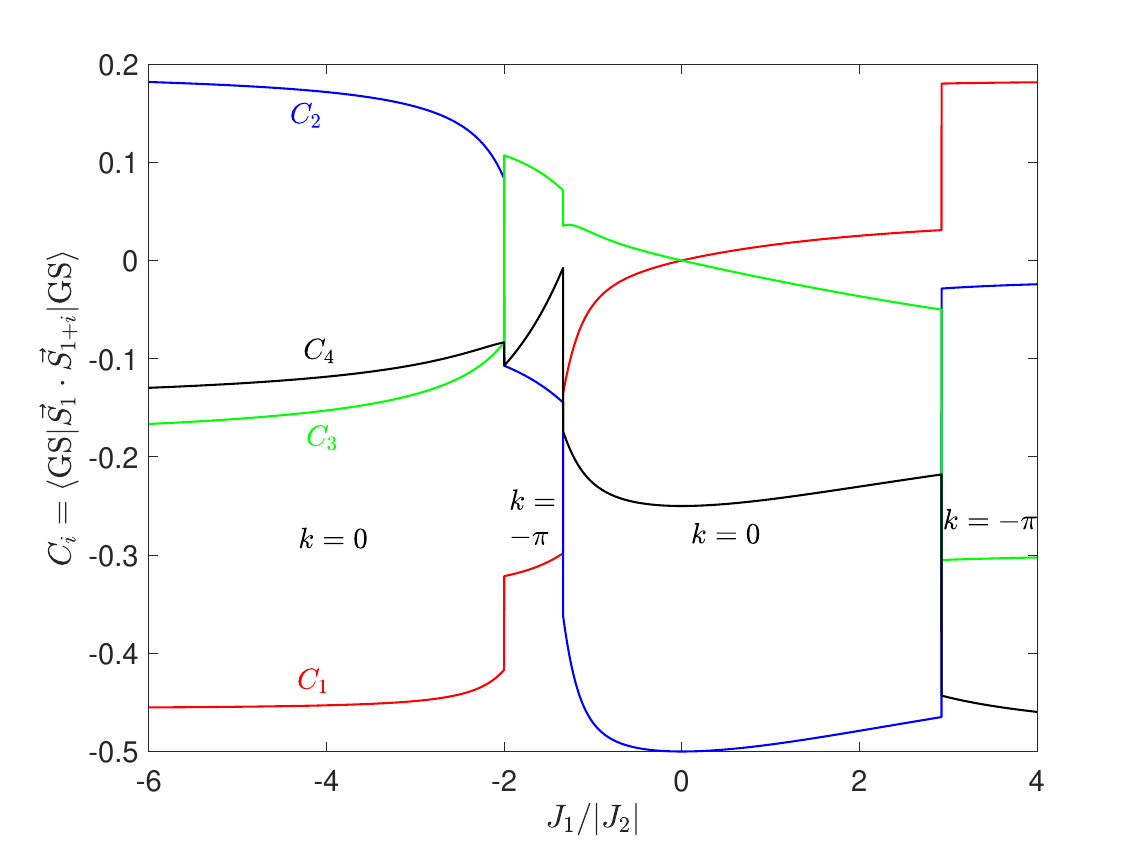}
\caption{Ground-state spin correlations $C_r=\langle \mathrm{GS}|\vec{S}_1\vec{S}_{1+r}|\mathrm{GS}\rangle$ ($r=1,2,3$, and $4$).}
\label{spincorr}
\end{figure}

 For completeness, we also plot the ground-state spin correlation functions $C_r= \langle\mathrm{GS}|\vec{S}_1\cdot\vec{S}_{1+r}|\mathrm{GS}\rangle=\frac{1}{N}\sum_i \langle\mathrm{GS}|\vec{S}_i\cdot\vec{S}_{i+r}|\mathrm{GS}\rangle$ in Fig.~\ref{spincorr}. These results are of course consistent with those obtained by exact diagonalization~\cite{TonegawaAFM,Tonegawa1989}. The behaviors of $C_r$ roughly reflect the evolution of the corresponding weights. For example, $|\chi_{10}(0)\rangle=|\xi_{2,2,2}(0)\rangle$ ($|\chi_1(-\pi)\rangle=|\xi_{1,1,1}(-\pi)\rangle$) has the largest weight in the region $J_1/|J_2|<-2$ ($2.924<J_1/|J_2|<4$), in accordance with the fact that $C_2$ ($C_1$) has large values in this region.  
\subsection{Exactly soluble points}
\par We mentioned that the two block Hamiltonians $h_0(-\pi)$ and $h_0(0)$ can be analytically solved for some particular values of $J_1/|J_2|$. We find by trials that $h_0(-\pi)$ [$h_0(0)$] has analytical solutions at $J_1/|J_2|=-6,-2,-2/3,0,2/3,2$, and $4$ ($J_1/|J_2|=-12,-6,-4,-2$, and $0$). Hence, both the MG ground states and the HKNN ground state can be analytically obtained:
\begin{eqnarray}
|\psi^{(\mathrm{e})}_{\mathrm{MG}}\rangle&=&\frac{\sqrt{2}}{3}[-|\xi_{1,2,2}(0)\rangle+|\xi_{1,2,3}(0)\rangle-|\xi_{1,3,2}(0)\rangle\nonumber\\
&&+|\xi_{1,3,1}(0)\rangle/\sqrt{2}+|\xi_{2,2,2}(0)\rangle],
\end{eqnarray} 
\begin{eqnarray}
|\psi^{(\mathrm{o})}_{\mathrm{MG}}\rangle&=&\sqrt{\frac{2}{7}}[-e^{-i\frac{\pi}{4}}|\xi_{1,2,2}(-\pi)\rangle+|\xi_{1,2,3}(-\pi)\rangle\nonumber\\
&&-e^{i\frac{\pi}{4}}|\xi_{1,3,2}(-\pi)\rangle+|\xi_{1,3,1}(-\pi)\rangle/\sqrt{2}],
\end{eqnarray} 
and 
\begin{eqnarray}\label{psiHKNN8}
|\psi_{\mathrm{HKNN}}\rangle&=&\frac{1}{\sqrt{62}}[6|\xi_{1,1,1}(-\pi)\rangle+3e^{i\frac{\pi}{4}}|\xi_{1,1,2}(-\pi)\rangle\nonumber\\
&&+3e^{-i\frac{\pi}{4}}|\xi_{1,1,4}(-\pi)\rangle+e^{-i\frac{\pi}{4}}|\xi_{1,2,2}(-\pi)\rangle\nonumber\\
&&+2|\xi_{1,2,3}(-\pi)\rangle+e^{i\frac{\pi}{4}}|\xi_{1,3,2}(-\pi)\rangle\nonumber\\
&&+\sqrt{2}|\xi_{1,3,1}(-\pi)\rangle].
\end{eqnarray} 
Note that the first five Bloch states $|\chi_1\rangle,\ldots,|\chi_5\rangle$ do not appear in either $|\psi^{(\mathrm{e})}_{\mathrm{MG}}\rangle$ or $|\psi^{(\mathrm{o})}_{\mathrm{MG}}\rangle$. From Eq.~(\ref{psiHKNN8}) we see again that the HKNN ground state $|\psi_{\mathrm{HKNN}}\rangle$ has $|\xi_{1,1,1}(-\pi)\rangle$ as its largest weighted component state. Thus, we conjecture that this property holds for arbitrary even number $N$, i.e., the Bloch state $|\xi_{1,1,\ldots,1}(-\pi)\rangle$ (there are $N/2-1$ $1$'s) should have the largest weight in $|\psi_{\mathrm{HKNN}}(N)\rangle$. 
\section{Conclusions}\label{SecV}
\par In this work, we use exact few-magnon Bloch states to block diagonalize a six-site anisotropic and an eight-site isotropic spin-1/2 $J_1$-$J_2$ ring. Using suitable good quantum numbers such as the number of magnons $n$, the wave number $k$, and the total angular momentum $l$, the total Hamiltonian is decomposed into smaller block Hamiltonians whose dimensions are at most four. As a byproduct, we realize a successful decomposition of the 14-dimensional zero-angular-momentum subspace of the eight-site ring, which was original considered by Majumdar and Ghosh~\cite{MG}. 
\par Even though the total angular momentum of the six-site anisotropic ring is not conserved, we find a subset of eigenstates that are simultaneous eigenstates of the Hamiltonian and the total angular momentum operator. The remaining discussions are mainly concerned with the isotropic case where the total angular momentum is conserved. For both $N=6$ and $8$, the full spectrum are obtained by solving the block Hamiltonians either analytically or numerically. In particular, we obtain explicit momentum-space expressions of the zero-energy state, the two MG ground states, and the HKNN ground state. We show the equivalence of the latter two with their real-space counterparts which are written in terms of dimer states. We also calculate the weights of Bloch states and spin-spin correlations in the ground state and the latter are consistent with previous works. We find that for both $N=6$ and $8$ the HKNN ground state has the Bloch state with $N/2$ successive down spins as its largest weighted component state. This suggest us to conjecture that for any even number $N$ the HKNN ground state may behave like a bound state with $N/2$ down spins binding together. 
\par The Bloch-state method used in this work can in principle be extended to larger $J_1$-$J_2$ rings, but more complicated calculations are needed. Nevertheless, the full spectrum and eigenstates obtained here for smaller rings may be helpful in the study of their dynamical or thermodynamical properties.\\ 
\\
\noindent{\bf Acknowledgments:}
We thank Dazhi Xu for useful discussions and critical reading of the manuscript. This work was supported by the Innovation Program for Quantum Science and Technology under Grant No. 2023ZD0300700 and by the National Key Research and Development Program of China under Grant No. 2021YFA1400803.
\\
\appendix
\section{Rules for constructing the Bloch Hamiltonians}\label{AppA}
\par We first consider the construction of the two-magnon Bloch Hamiltonians. Starting with the real-space state $|1,1+r\rangle=S^-_1S^-_{1+r}|F\rangle$ and applying the operator $X_m\equiv \sum^N_{j=1}(S^+_jS^-_{j+m}+S^-_jS^+_{j+m})$
to it, we will get some other states characterized by $r'$, $T^{j_1}|1,1+r'\rangle,T^{j_2}|1,1+r'\rangle$, etc. Then, the matrix element turns out to be~\cite{PRB2022}
\begin{eqnarray}
\langle \xi_{r'}(k)|X_m|\xi_r(k)\rangle=e^{i\frac{k}{2}(r-r')}\sum_ie^{-ikj_i}.
\end{eqnarray}
For example, from $X_1|\xi_1\rangle=T^{-1}|\xi_2\rangle+|\xi_2\rangle$ we get $\langle \xi_{2}(k)|X_1|\xi_1(k)\rangle=e^{-i\frac{k}{2}}(1+e^{ik})=2\cos \frac{k}{2}$.
\par Similarly, in the three-magnon sector we start with the state $|1,1+r_1,1+r_1+r_2\rangle=S^-_1S^-_{1+r_1}S^-_{1+r_1+r_2}|F\rangle$ and apply $X_m$ to it to get some other states characterized by $(r'_1,r'_2)$, $T^{j_1}|1,1+r'_1,1+r'_1+r'_2\rangle,T^{j_2}|1,1+r'_1,1+r'_1+r'_2\rangle$, etc.
The matrix element is then given by ($z=e^{-ik/3}$)
\begin{eqnarray}
\langle \xi_{r'_1,r'_2}(k)|X_m|\xi_{r_1,r_2}(k)\rangle=z^{2(r'_1-r_1)+(r'_2-r_2)}\sum_i z^{3j_i}.
\end{eqnarray}
For example, from $X_1|\xi_{1,1}\rangle=T|\xi_{1,N-3}\rangle+|\xi_{1,2}\rangle$ we have $\langle \xi_{1,N-3}(k)|X_1|\xi_{1,1}(k)\rangle=z^{N-4} z^{3}=z^{N-1}$ and $\langle \xi_{1,2}(k)|X_1|\xi_{1,1}(k)\rangle=z$.
\par Finally, in the four-magnon sector we start with the state $|1,1+r_1,1+r_1+r_2,1+r_1+r_2+r_3\rangle$ and apply $X_m$ to it to get some other states characterized by $(r'_1,r'_2,r'_3)$: $T^{j_1}|1,1+r'_1,1+r'_1+r'_2,1+r'_1+r'_2+r'_3\rangle$, $\cdots$, the matrix element is then given by ($\tilde{z}=e^{-ik/4}$)
\begin{eqnarray}
\langle \xi_{r'_1,r'_2,r'_3}(k)|X_m|\xi_{r_1,r_2,r_3}(k)\rangle&=&\tilde{z}^{3(r'_1-r_1)+2(r'_2-r_2)+(r'_3-r_3)}\nonumber\\
&&\times\sum_i \tilde{z}^{4j_i}.
\end{eqnarray}
\section{Matrix representation of $\vec{L}^2$}\label{AppB}
The total angular momentum $\vec{L}^2$ for $N=6$ can be rewritten as
\begin{eqnarray}\label{L2}
\vec{L}^2&=&\frac{9}{2}+H^{(\mathrm{isp})}|_{J_1=J_2=-2}\nonumber\\ &&+2(\vec{S}_1\cdot\vec{S}_4+\vec{S}_2\cdot\vec{S}_5+\vec{S}_3\cdot\vec{S}_6).
\end{eqnarray}
The second line in $\vec{L}^2$ corresponds to a next-next-nearest-neighbor interaction. Using the rules provided in Appendix~\ref{AppA}, we obtain  
\begin{eqnarray}\label{L2matrix0}
\vec{L}^2_0&=& 3(3+1)=12,
\end{eqnarray}
\begin{eqnarray}\label{L2matrix1}
\vec{L}^2_1(k\in K_1)&=&7+2(\cos k+\cos 2k)+\cos 3k,
\end{eqnarray}
\begin{eqnarray}\label{L2matrix2}
\vec{L}^2_2\left(\pm\frac{2\pi}{3}\right)&=&  \left(
                \begin{array}{cccc}
                  3     & -1  & -\sqrt{2}    \\
                  -1 &  3   &  \sqrt{2}     \\
                  -\sqrt{2}  &  \sqrt{2}  &  4     \\
                \end{array}
              \right),\nonumber\\
              \vec{L}^2_2\left(0\right)&=& 2 \left(
                \begin{array}{cccc}
                  3    & 2 & \sqrt{2}   \\
                  2 &  3  &  \sqrt{2}    \\
                  \sqrt{2} &  \sqrt{2}&  2     \\
                \end{array}
              \right),\nonumber\\
\vec{L}^2_2(-\pi)&=& \left(
                \begin{array}{cccc}
                  2 & 0     \\
                0  & 6    \\
                \end{array}
              \right),\nonumber\\
               \vec{L}^2_2\left(\pm\frac{\pi}{3}\right)&=&\left(
                \begin{array}{cccc}
                  5 & \sqrt{3}     \\
                 \sqrt{3}   & 3    \\
                \end{array}
              \right),
\end{eqnarray}
\begin{eqnarray}\label{L2matrix3}
\vec{L}^2_3(-\pi)&=&  \left(
                \begin{array}{cccc}
                    1  & -e^{-i\frac{\pi}{3}} & -e^{i\frac{\pi}{3}} &  -\sqrt{3}  \\
                   -e^{i\frac{\pi}{3}} &  2  & 0 & \sqrt{3}e^{i\frac{\pi}{3}} \\
                  -e^{-i\frac{\pi}{3}} & 0 &  2  & \sqrt{3}e^{-i\frac{\pi}{3}} \\
                  -\sqrt{3}  & \sqrt{3}e^{-i\frac{\pi}{3}}  & \sqrt{3}e^{i\frac{\pi}{3}} & 3 \\
                \end{array}
              \right),\nonumber\\
              \vec{L}^2_3(0)&=& \left(
                \begin{array}{cccc}
                    5    & 3 & 3 &  \sqrt{3}  \\
                  3 &  4  & 4 & \sqrt{3}  \\
                  3 & 4 &  4  & \sqrt{3}  \\
                  \sqrt{3} & \sqrt{3}   & \sqrt{3}  & 3 \\
                \end{array}
              \right), \nonumber\\
\vec{L}^2_3\left(-\frac{2\pi}{3}\right)&=& 2 \left(
                \begin{array}{cccc}
                  1    & 0 & 0   \\
                  0 &  2  &  e^{i\frac{\pi}{9}}    \\
                  0 &  e^{-i\frac{\pi}{9}} &  2     \\
                \end{array}
              \right),\nonumber\\
              \vec{L}^2_3\left(-\frac{\pi}{3}\right)&=& 2 \left(
                \begin{array}{cccc}
                  2   & e^{-i\frac{\pi}{9}} & -e^{i\frac{4\pi}{9}}   \\
                  e^{i\frac{\pi}{9}} &  1  &  0    \\
                  -e^{-i\frac{4\pi}{9}} &  0 &  1     \\
                \end{array}
              \right), \nonumber\\
\vec{L}^2_3\left(\frac{2\pi}{3}\right)&=&\vec{L}^{2*}_3\left(-\frac{2\pi}{3}\right),\nonumber\\
\vec{L}^2_3\left(\frac{\pi}{3}\right)&=&\vec{L}^{2*}_3\left(-\frac{\pi}{3}\right).
\end{eqnarray}
\section{Solutions of the two- and three-magnon Bloch Hamiltonians for $N=6$}\label{App23}
\subsection{Two-magnon sector}
\par For $k=-\pi\in K'_2$, both $H_2(-\pi)$ and $\vec{L}^2_2(-\pi)$ are diagonal, so the Bloch state $|\xi_1(-\pi)\rangle$ [$|\xi_2(-\pi)\rangle$] is itself a simultaneous eigenstate of $H$ and $\vec{L}^2$ with eigenenergy $E^{(1)}_2(-\pi)=-J_-+J_2$ [$E^{(2)}_2(-\pi)=J_--J_2$] and angular momentum $l=1$ ($l=2$). Therefore, $|\xi_1(-\pi)\rangle$ and $|\xi_2(-\pi)\rangle$ are two type-HL eigenstates.
\par For $k=-\pi/3\in K'_2$, diagonalization of the $2\times 2$ matrix $H_2(-\pi/3)$ gives $E^{(1)}_2(-\pi/3)=-B_1/2$ and $E^{(2)}_2(-\pi/3)=B_1/2$, where $B_1\equiv \sqrt{4J_-(J_-+J_2)+3J^2_1+J^2_2}$. The corresponding (unnormalized) eigenstates are
\begin{eqnarray}\label{psi21pi3}
|\psi^{(1)}_2(-\pi/3)\rangle&=&\left(
   \begin{array}{c}
     J_2+2J_-+B_1 \\
     \sqrt{3}J_1 \\
   \end{array}
 \right),\nonumber\\
 |\psi^{(2)}_2(-\pi/3)\rangle&=&\left(
   \begin{array}{c}
     J_2+2J_--B_1 \\
     \sqrt{3}J_1 \\
   \end{array}
 \right).
\end{eqnarray}
However, $|\psi^{(1)}_2(-\pi/3)\rangle$ and $|\psi^{(2)}_2(-\pi/3)\rangle$ are not eigenstates of $\vec{L}^2_2(-\pi/3)$ unless $\Delta_1=\Delta_2=1$.
\par For $k=0$ and $-2\pi/3$ belonging to $K_2$, Mathematica fails to give closed-form expressions for the eigenvalues and eigenvectors of the $3\times 3$ matrices $H_2(0)$ and $H_2(-2\pi/3)$. Nevertheless, for $H^{(\mathrm{isp})}_2(0)$ and $H^{(\mathrm{isp})}_2(-2\pi/3)$ we obtain
\begin{eqnarray}
E_{2,3}(0)&=&-3(J_1+J_2)/2,\nonumber\\
E^{(1)}_{2,1}(0)&=&J_1-B_0/2,~E^{(2)}_{2,1}(0)=J_1+B_0/2,
\end{eqnarray}
with eigenstates ($\mathrm{T}$ denotes matrix transpose)
\begin{eqnarray}\label{psi20}
|\psi_{2,3}(0)\rangle&=&\left(
                                \begin{array}{ccc}
                                  \sqrt{2} & \sqrt{2} & 1 \\
                                \end{array}
                              \right)^{\mathrm{T}}/\sqrt{5},\nonumber\\
|\psi^{(1)}_{2,1}(0)\rangle&=&\left(
   \begin{array}{c}
     -J^2_1-4J_1J_2+3J^2_2-(J_1+J_2)B_0 \\
     J_1(-J_1+5J_2+B_0)\\
     \sqrt{2}(2J^2_1-J_1J_2-3J^2_2+J_2B_0) \\
   \end{array}
 \right),\nonumber\\
 |\psi^{(2)}_{2,1}(0)\rangle&=& \left(
   \begin{array}{c}
     -J^2_1-4J_1J_2+3J^2_2+(J_1+J_2)B_0 \\
     J_1(-J_1+5J_2-B_0)\\
     \sqrt{2}(2J^2_1-J_1J_2-3J^2_2-J_2B_0) \\
   \end{array}
 \right),\nonumber\\
\end{eqnarray}
and
\begin{eqnarray}
E_{2,2}(-2\pi/3)&=&0,\nonumber\\
E^{(1)}_{2,1}(-2\pi/3)&=&(J_1+3J_2-B_2)/4,\nonumber\\
E^{(2)}_{2,1}(-2\pi/3)&=&(J_1+3J_2+B_2)/4,
\end{eqnarray}
with eigenstates
\begin{eqnarray}\label{psi22pi3}
|\psi_{2,2}(-2\pi/3)\rangle&=& \left(
                                \begin{array}{ccc}
                                  -1 & 1 & \sqrt{2} \\
                                \end{array}
                              \right)^{\mathrm{T}}/2,\nonumber\\
|\psi^{(1)}_{2,1}(-2\pi/3)\rangle&=&\left(
   \begin{array}{c}
     5J^2_1-7J_1J_2+6J^2_2+(J_1-2J_2)B_2 \\
     J_1(J_1-5J_2+B_2)\\
     \sqrt{2}(2J^2_1-J_1J_2+3J^2_2-J_2B_2) \\
   \end{array}
 \right),\nonumber\\
|\psi^{(2)}_{2,1}(-2\pi/3)\rangle&=&\left(
   \begin{array}{c}
     5J^2_1-7J_1J_2+6J^2_2-(J_1-2J_2)B_2 \\
     J_1(J_1-5J_2-B_2)\\
     \sqrt{2}(2J^2_1-J_1J_2+3J^2_2+J_2B_2) \\
   \end{array}
 \right),\nonumber\\
\end{eqnarray}
where $B_0\equiv \sqrt{5J^2_1-10J_1 J_2+9J^2_2}$ and $B_2=\sqrt{17J^2_1-10J_1J_2+9J^2_2}$. Note that the state $|\psi_{2,3}(0)\rangle$ [$|\psi_{2,2}(-2\pi/3)\rangle$] is degenerate with the one-magnon state $|\xi(0)\rangle$ [$|\xi(-2\pi/3)\rangle$] and the two are connected by $\mathcal{L}_0$, so their wave functions are independent of $J_1$ and $J_2$. 
\subsection{Three-magnon sector}
\par For $k=-\pi\in K_3$, Mathematica can only output a single closed-form eigenvalue of $H_3(-\pi)$, $E_3(-\pi)=J_+-J_1+J_2$. The corresponding eigenstate $|\psi_3(-\pi)\rangle=\left(
                        \begin{array}{cccc}
                          0 & e^{-i\frac{\pi}{3}} & 1 & 0 \\
                        \end{array}
                      \right)^{\mathrm{T}}/\sqrt{2}$ is a type-HL state belonging to $l=1$. For $\Delta_1=\Delta_2=1$, the remaining three eigenvalues are
\begin{eqnarray}
E_{3,2}(-\pi)&=&\frac{1}{2}(J_1-3J_2),\nonumber\\
E^{(1)}_{3,0}(-\pi)&=&J_1-\frac{A_3}{2},~E^{(2)}_{3,0}(-\pi)=J_1+\frac{A_3}{2},
\end{eqnarray}
with eigenvectors
\begin{eqnarray}\label{psi3pi}
|\psi_{3,2}(-\pi)\rangle&=&\left(
                             \begin{array}{cccc}
                               -1 & e^{i\frac{\pi}{3}} & e^{-i\frac{\pi}{3}} & \sqrt{3} \\
                             \end{array}
                           \right)^{\mathrm{T}}/\sqrt{6},\nonumber\\
|\psi^{(1)}_{3,0}(-\pi)\rangle&=&       \left(
  \begin{array}{c}
    -2J_1(J_1+3J_2+A_3)\\
   (J_1-3J_2+A_3)(3J_1-3J_2-A_3)e^{i\frac{\pi}{3}}\\
   -2J_1 (5J_1-3J_2-A_3)e^{-i\frac{\pi}{3}} \\
   2 \sqrt{3}J_1(3J_1-3J_2-A_3) \\
  \end{array}
\right),\nonumber\\
  |\psi^{(2)}_{3,0}(-\pi)\rangle&=&          \left(
  \begin{array}{c}
    -2J_1(J_1+3J_2-A_3) \\
   (J_1-3J_2-A_3)(3J_1-3J_2+A_3) e^{i\frac{\pi}{3}}\\
    -2J_1(5J_1-3J_2+A_3)e^{-i\frac{\pi}{3}} \\
    2\sqrt{3}J_1(3J_1-3J_2+A_3) \\
  \end{array}
\right),\nonumber\\
\end{eqnarray}
where $A_3\equiv \sqrt{13J^2_1-18 J_1J_2+9J^2_2}$.
\par Similarly, for $k=0\in K_3$, we have a single type-HL eigenstate $|\psi_{3}(0)\rangle=\left(
                                                                                             \begin{array}{cccc}
                                                                                               0 & -1 & 1 & 0 \\
                                                                                             \end{array}
                                                                                           \right)^{\mathrm{T}}/\sqrt{2}$, which belongs to $l=0$ and has energy $E_{3}(0)=J_++J_1+J_2$. The remaining three eigenenergies for $\Delta_1=\Delta_2=1$ read
\begin{eqnarray}
E_{3,3}(0)&=&-\frac{3}{2}(J_1+J_2),\nonumber\\
E^{(1)}_{3,0}(-\pi)&=&J_1-\frac{A_0}{2},~E^{(2)}_{3,0}(-\pi)=J_1+\frac{A_0}{2},
\end{eqnarray}
with eigenvectors
\begin{eqnarray}
|\psi_{3,3}(0)\rangle&=&\left(
                          \begin{array}{cccc}
                            \sqrt{3} & \sqrt{3} & \sqrt{3} & 1 \\
                          \end{array}
                        \right)^{\mathrm{T}}/\sqrt{10},\nonumber\\
|\psi^{(1)}_{3,1}(0)\rangle&=&   \left(
   \begin{array}{c}
     -2(2J_1-3J_2+A_0) \\
     J_1-3J_2+A_0 \\
     J_1-3J_2+A_0\\
     2\sqrt{3}J_1 \\
   \end{array}
 \right),\nonumber\\
 |\psi^{(2)}_{3,1}(0)\rangle&=&   \left(
   \begin{array}{c}
     -2(2J_1-3J_2-A_0) \\
     J_1-3J_2-A_0 \\
     J_1-3J_2-A_0 \\
     2\sqrt{3}J_1 \\
   \end{array}
 \right),
\end{eqnarray}
where $A_0\equiv \sqrt{5J^2_1-10J_1J_2+9J^2_2}$.
\par The eigenenergies of $H_3(-2\pi/3)$ are
\begin{eqnarray}
E^{(1)}_{3}(-2\pi/3)&=&J_+-\frac{1}{2}(J_1+J_2),\nonumber\\
E^{(2)}_{3}(-2\pi/3)&=&\frac{1}{4}[(1+2\Delta_2)J_2+J_1-A_2],\nonumber\\
E^{(3)}_{3}(-2\pi/3)&=&\frac{1}{4}[(1+2\Delta_2)J_2+J_1+A_2],
\end{eqnarray}
with eigenstates
 \begin{eqnarray}\label{psi32pi3}
|\psi^{(1)}_{3}(-2\pi/3)\rangle&=&\left(
                                    \begin{array}{ccc}
                                      0 & e^{i\frac{\pi}{9}} & 1 \\
                                    \end{array}
                                  \right)^{\mathrm{T}}/\sqrt{2} ,\nonumber\\
|\psi^{(2)}_{3}(-2\pi/3)\rangle&=&   \left(
   \begin{array}{c}
    [(1+2\Delta_1)J_1+J_2+A_2]e^{-i\frac{\pi}{9}} \\
   2(J_1-J_2) e^{i\frac{\pi}{9}} \\
     -2(J_1-J_2) \\
   \end{array}
 \right),\nonumber\\
 |\psi^{(3)}_{3}(-2\pi/3)\rangle&=&   \left(
   \begin{array}{c}
    [(1+2\Delta_1)J_1+J_2-A_2]e^{-i\frac{\pi}{9}} \\
   2(J_1-J_2) e^{i\frac{\pi}{9}} \\
     -2(J_1-J_2) \\
   \end{array}
 \right),\nonumber\\
\end{eqnarray}
 where $A_2\equiv \sqrt{(4\Delta^2_1+4\Delta_1+9)J^2_1+2(2\Delta_1-7)J_1 J_2+9J^2_2}$. These three states are all type-HL eigenstates and belong to
$l=2,1$, and $1$, respectively. 
\par Though Mathematica fails to provide analytical expressions for the eigenenergies of $H_3(-\pi/3)$, we checked that they are indeed given by
\begin{eqnarray}
E^{(1)}_{3}(-\pi/3)&=&J_++\frac{1}{2}(J_1-J_2),\nonumber\\
E^{(2)}_{3}(-\pi/3)&=&\frac{1}{4}[(1+2\Delta_2)J_2-J_1-A_1],\nonumber\\
E^{(3)}_{3}(-\pi/3)&=&\frac{1}{4}[(1+2\Delta_2)J_2-J_1+A_1],
\end{eqnarray}
where $A_1\equiv \sqrt{(4\Delta^2_1-4\Delta_1+9)J^2_1+2(2\Delta_1+7)J_1 J_2+9J^2_2}$. The eigenstate corresponding to $E^{(1)}_{3}(-\pi/3)$ reads $|\psi^{(1)}_3(-\pi/3)\rangle=\left(
                                    \begin{array}{ccc}
                                      0 & e^{i\frac{5\pi}{9}} & 1 \\
                                    \end{array}
                                  \right)^{\mathrm{T}}/\sqrt{2}
$, which is a type-HL state and belongs to $l=1$. The remaining two eigenstates $|\psi^{(2)}_3(-\pi/3)\rangle$ and $|\psi^{(3)}_3(-\pi/3)\rangle$ generally  do not have definite angular momentum since $\vec{L}^2$ is nondegenerate in the $k=-\pi/3$ sector (see Table I). At the isotropic point, we have $A^2_1=9(J_1+J_2)^2$, so that the three eigenenergies become $J_1$, $-J_1$, and $(J_1+3J_2)/2$, belonging to $l=1$, $2$, and $0$, respectively.

\section{Expressing the dimer states in terms of real-space three-magnon basis states}\label{AppC}
\par The $15$ dimer states appearing in $|\psi_{\mathrm{HKNN}}\rangle$ can be expressed in terms of the $20$ real-space three-magnon basis states as follows by directly expanding the singlets:
\begin{widetext}
\begin{eqnarray}\label{dimermagnon}
~[12][34][56]&=&(T^{-1}+T+T^3)(|1,2,4\rangle-|1,2,5\rangle) +(T-1)|1,3,5\rangle,\nonumber\\
~[12][35][46]&=&(T^{-2}+T^{-1}+T^3)|1,2,4\rangle -(T+T^2+T^3)|1,2,5\rangle+(T-T^{-2})|1,2,3\rangle,\nonumber\\
~[12][36][45]&=&(T^{-2}-T)(|1,2,4\rangle-|1,2,3\rangle) +(T^{-1}-T^2)|1,2,5\rangle+(1-T)|1,3,5\rangle,\nonumber\\
~[13][24][56]&=&(T+T^2+T^3)|1,2,4\rangle -(T^{-1}+1+T)|1,2,5\rangle+(T^{-1}-T^2)|1,2,3\rangle,\nonumber\\
~[13][25][46]&=&(T^{-2}-T)|1,2,5\rangle+(T^{3}-1)|1,2,4\rangle +(T+T^{-1}-T^{-2}-T^2)|1,2,3\rangle,\nonumber\\
~[13][26][45]&=&(T-T^{-2})|1,2,3\rangle-(1+T+T^2)|1,2,4\rangle +(T^{-1}+1+T^{-2})|1,2,5\rangle,\nonumber\\
~[14][23][56]&=&(T^2-T^{-1})(|1,2,4\rangle-|1,2,3\rangle) +(T^3-1)|1,2,5\rangle+(1-T)|1,3,5\rangle,\nonumber\\
~[14][25][36]&=&(T^3-T^{-2}-T^2+T+T^{-1}-1)|1,2,3\rangle +(1-T)|1,3,5\rangle,\nonumber\\
~[14][26][35]&=&(1-T^3)|1,2,5\rangle+(T^{-1}-T^2)|1,2,4\rangle +(T^3-1-T^{-2}+T)|1,2,3\rangle,\nonumber\\
~[15][23][46]&=&(T^{-2}+T^3+T^2)|1,2,5\rangle+(T^{-1}-T^2)|1,2,3\rangle -(1+T^{-1}+T^{-2})|1,2,4\rangle,\nonumber\\
~[15][24][36]&=&(T^2-T^{-1})|1,2,5\rangle+(T-T^{-2})|1,2,4\rangle +(T^3-1-T^2+T^{-1})|1,2,3\rangle,\nonumber\\
~[15][26][34]&=&(T^3-1)|1,2,3\rangle-(T^{-2}+T^{-1}+T^3)|1,2,5\rangle +(T^{-1}+1+T)|1,2,4\rangle,\nonumber\\
~[16][23][45]&=&(T^{-2}+1+T^2)(|1,2,5\rangle-|1,2,4\rangle) +(T-1)|1,3,5\rangle,\nonumber\\
~[16][24][35]&=&(T^3-1)|1,2,3\rangle+(1+T+T^2)|1,2,5\rangle -(T^{-2}+T^2+T^3)|1,2,4\rangle,\nonumber\\
~[16][25][34]&=&(T^3-1)(|1,2,3\rangle-|1,2,4\rangle) +(T-T^{-2})|1,2,5\rangle+(1-T)|1,3,5\rangle.
\end{eqnarray}
\end{widetext}

\end{document}